\documentclass[a4paper,onecolumn,11pt,unpublished]{quantumarticle}
\pdfoutput=1
\usepackage[utf8]{inputenc}
\usepackage[english]{babel}
\usepackage[T1]{fontenc}
\usepackage{amsmath} 
\usepackage{amsfonts}
\usepackage{amssymb}
\usepackage{amsthm}
\usepackage{bm} 
\usepackage{bbm}
\usepackage{bbold}
\usepackage{dsfont}
\usepackage{mathrsfs}
\usepackage{subcaption}
\usepackage{tabularx}
\usepackage{cellspace}
\usepackage{dcolumn} 
\usepackage{graphicx} 
\usepackage{svg}
\usepackage{mwe}
\usepackage{tikz}
\usepackage{tikz-cd}
\usetikzlibrary{quantikz2}
\usepackage{xcolor}
\usepackage{hyperref}
\usepackage[capitalise]{cleveref}
\hypersetup{citecolor=blue}
\usepackage[numbers]{natbib}

\newcommand{\hc}{\mathrm{h.c.}}
\newcommand{\rre}{\mathrm{Re}}
\newcommand{\rim}{\mathrm{Im}}

\newcommand{\im}{i}

\newcommand{\Tr}{\op{Tr}}
\newcommand{\id}{\mathbb{1}}

\theoremstyle{definition}
\newtheorem{definition}{Definition}
\newtheorem{proposition}{Proposition}

\newcommand{\red}[1]{#1}

\newcommand{\steinergauss}{\textsc{steiner-gauss}\xspace}
\newcommand{\steinerup}{\textsc{steiner-up}\xspace}
\newcommand{\steinerdown}{\textsc{steiner-down}\xspace}

\newcommand{\cxg}[1]{\gate[style={minimum height=0.5cm}]{\scriptstyle \mathrm{X}^{#1}} }

\newcommand{\xonetwo}{\gate{\scriptstyle X^{(12)}}}

\newcommand{\grz}[1]{\gate{\scriptstyle R_{#1}}}

\newcommand{\cx}{\ensuremath{\mathrm{CX}}\xspace}
\newcommand{\cnot}{\ensuremath{\mathrm{CNOT}}\xspace}
\newcommand{\swp}{\ensuremath{\mathrm{SWAP}}\xspace}
\newcommand{\rs}{R_{0}(\theta)}

\newcommand\scalemath[2]{\scalebox{#1}{\mbox{\ensuremath{\displaystyle #2}}}}

\newcommand{\op}[1]{\operatorname{#1}}
\newcommand{\wpg}{\ensuremath{G(\theta, s, \bm{k})}\xspace}

\tikzset{
  gateV/.style={
    draw,
    inner sep=8pt,
  },
  gateO/.style={
    draw,
    circle,
    minimum width=1.0em,
    inner sep=2pt
  },
  gateQ/.style={
    draw,
    inner sep=8pt,
    inner ysep=0pt,
  }
}

\DeclareExpandableDocumentCommand{\gateO}{O{}{m}}{|[gateO,#1]| {#2} \qw}

\begin{document}

\title{Decomposition of multi-qutrit gates generated by Weyl-Heisenberg strings}

\author{Daniele Trisciani}
\affiliation{QTF Centre of Excellence, Department of Physics, University of Helsinki, Helsinki, Finland}
\author{Marco Cattaneo}
\affiliation{QTF Centre of Excellence, Department of Physics, University of Helsinki, Helsinki, Finland}
\author{Zolt\'an Zimbor\'as}
\affiliation{QTF Centre of Excellence, Department of Physics, University of Helsinki, Helsinki, Finland}
\affiliation{Wigner Research Centre for Physics, Budapest, Hungary}
\affiliation{Algorithmiq Ltd, Kanavakatu 3C 00160 Helsinki, Finland}

\maketitle
\begin{abstract}
  Decomposing unitary operations into native gates is an essential step for implementing quantum algorithms. Motivated by the growing interest in qutrit-based quantum computing, we introduce a technique to exactly decompose any diagonal qutrit unitary into single-qutrit rotations and two-qutrit CX gates, thus enabling the decomposition of the exponential of any sum of commuting Hermitian Weyl-Heisenberg strings. To achieve this we introduce the \emph{Weyl phase gadget}, that, with similar properties as its qubit counterpart, the Pauli phase gadget, serves as intermediate step in the decomposition. As a practical application, we use our method to decompose the layers of the quantum approximate optimization algorithm for qutrit-based implementations of the graph $k$-coloring problem. For values of $k$ well-suited to qutrit architectures (e.g., $k=3$ or in general $k=3^n$), our approach yields significantly shallower circuits compared to qubit-based implementations, an advantage that grows with problem size, while also requiring a smaller total Hilbert space dimension. Finally, we address the routing challenge in qutrit architectures that arises due to the limited connectivity of the devices. In particular, we generalize the \steinergauss method, originally developed to reduce CNOT counts in qubit circuits, to optimize gate routing in qutrit-based systems.
\end{abstract}

\section{Introduction}

The promise of surpassing the capabilities of classical computing with quantum computers has driven rapid advancements in both quantum hardware and software. However, today’s quantum hardware, commonly referred to as Noisy Intermediate Scale Quantum (NISQ) devices~\cite{preskill2018quantum}, remain limited in several key aspects, including qubit count, coherence time, and error rates. These latter two impose strict constraints on circuit depth, posing challenges for realizing practical quantum applications~\cite{zimboras2025myths}. To address this, advances are needed not only in algorithm design but also in the efficient implementation of circuits that align with hardware capabilities. In this effort, gate decomposition plays an essential role as it bridges the gap between high-level quantum algorithms and the native gate sets of actual quantum processors.

Many physical systems naturally possess three energy levels that can be efficiently manipulated, which has led to growing interest in developing qutrit-based quantum computers~\cite{vepsalainen2019superadiabatic, Lu_qudits_2019, blok2021quantum, Morvan_2021, Goss_2022, chi2022, goss2024extending, patel2025harrow}. This trend, in turn, is driving theoretical research toward implementing general quantum algorithms using qutrits. On one front, existing qubit-based algorithms are being translated into qutrit-based counterparts~\cite{Bocharov_2017, wang2020, ivanov2021, roy2022}. On another front, researchers are exploring qutrit-native problems, such as graph 3-coloring or quantum simulation of spin-1 systems, where qutrits offer a more natural framework~\cite{bottrill2023, karacsony2024, kumaran2024, ogunkoya2024, heimsoth2026}. This trend is driving the development of new methods for qutrit unitary synthesis~\cite{li_2026, yeh2022constructing, mato_ringbauer_2023, jiang_2023, van_de_Wetering_2023}. Mixed qubit-qutrit solutions are also being explored~\cite{B_kkegaard_2019,Gokhale_2019,Tomesh_2024,mato2023}.

In this work, we develop new gate decomposition schemes for qutrit-based quantum computers. In particular, we introduce a scheme to exactly decompose any diagonal qutrit unitary into CX and diagonal single-qutrit gates. We also introduce a qutrit generalization of the Pauli phase gadget, the \emph{Weyl phase gadget}, which serves as intermediate step in the decomposition. The decomposition is optimal and, in principle, our method extends to arbitrary multi-qutrit gates, provided they are appropriately Trotterized~\cite{nielsenchuang}. As a practical application, we apply this decomposition to the qutrit-based quantum approximate optimization algorithm~\cite{farhi2014} focusing on the graph $k$-coloring problem. Finally, we derive an upper bound on the number of \cx gates required for the implementation of a generic diagonal qutrit unitary.

For the sake of clarity, we point out that our aim is not to decompose qutrit gates into qubit native gates; rather, we assume to work with a native set of qutrit gates, including the CX gate, available on the hardware (e.g., exploiting the third level of a transmon), so we need not concern ourselves with the additional decomposition of qutrit gates into qubit gates.

As many quantum computer platforms operate with restricted connectivity, various connectivity-aware decompositions have been developed for qubit systems~\cite{zulehner2018, herbert2019, kissinger2019,nash2020}. In particular, for decomposing exponentials of Pauli strings on devices with limited connectivity, the so-called \steinergauss method could be applied~\cite{kissinger2019, nash2020}, which leverages the solutions of the Steiner tree problem to optimize the gate decompositions. In this work, we generalize the \steinergauss method for the compilation of circuits on qutrit-based architectures.

The paper is structured as follows: Section \ref{sec:qubit_to_qutrits} introduces the elementary gates used for qutrit quantum computation, which form the basis for the decompositions discussed later. Section \ref{sec:pauli_decomp} provides a review of the decomposition of Pauli phase gadgets, while in Section \ref{sec:qutrit_decomp} we introduce the Weyl phase gadget, which serves for the decomposition of any diagonal qutrit unitary.
Section \ref{sec:decomp_application} applies this decomposition technique to instances of the graph $k$-coloring problem within the QAOA framework. The focus then shifts to Section \ref{sec:restrictedTop}, which addresses restricted topologies in qutrit systems. In particular, subsection \ref{sec:steiner_review} review the \steinergauss algorithm, and subsection \ref{sec:ternary_routing} extends this method to qutrit architectures by introducing the ternary parity map.

\section{Background} 

\subsection{From qubits to qutrits} \label{sec:qubit_to_qutrits}

Quantum computation traditionally relies on the qubit, the quantum analog of the classical bit. A qubit is a two-level quantum system with computational basis states denoted by $\ket{0}$ and $\ket{1}$ and arbitrary pure qubit states expressed as the superposition of these.
As the field matures, increasing attention is being given to qudits, i.e., quantum systems with $d>2$ levels. In particular, the qutrit, a three-level system with basis states $\ket{0}$, $\ket{1}$ and $\ket{2}$ offers a natural next step. A general pure qutrit state takes the form 
$\ket{\psi}=\alpha\ket{0} + \beta\ket{1} + \gamma\ket{2}$ with 
$|\alpha|^2 + |\beta|^2 + |\gamma|^2 =1$.

The Pauli matrices $\sigma^x$, $\sigma^y$, and $\sigma^z$, together with the identity, 
form a complete basis for the space of operators acting on the qubit Hilbert space. This basis is particularly convenient due to the favorable properties of its non-identity elements: they are traceless, self-adjoint, unitary, mutually orthogonal under the Hilbert-Schmidt inner product and satisfy anticommutation relations. Generalizing this structure to qutrit systems has led to several proposed operator bases. The two most widely used, the Gell-Mann matrices~\cite{gellmann_1962} and the Weyl-Heisenberg operators~\cite{Bennett_1993, Bertlmann_2008}, each offer distinct structural advantages depending on the task.

The Gell-Mann matrices are Hermitian and mutually orthogonal operators but are not, in general, unitary. By contrast, Weyl-Heisenberg operators are unitary and traceless, but not Hermitian. In this work, we adopt the latter. The Weyl-Heisenberg operators $X$ and $Z$ act on the qutrit computational states as
\begin{equation} \label{eq:qutritXZ}
  X\ket{j} = \ket{(j+1) \bmod 3} \quad \text{and} \quad
  Z\ket{j} = \omega^j\ket{j},
\end{equation}
for $\omega=e^{i \frac{2\pi}{3}}$.
Moreover, they satisfy the noncommutative relation $ZX=\omega XZ$.

\begin{definition} \label{def:hadamard}
  The qutrit Hadamard gate $H$ acts on the computational state as:
  \begin{equation} \label{eq:thadamard}
    H \ket{j} = \frac{1}{\sqrt{3}} \left(\ket{0} + \omega^{j} \ket{1} + \omega^{2j} \ket{2}\right).
  \end{equation}
\end{definition}
This gate transforms computational basis states of qutrits into a uniform superposition of the basis states. The qutrit Hadamard conjugates the Weyl-Heisenberg operators as $HZH^{\dagger}=X$, and satisfies $H^{\dagger}=H^3$ and $H^2 = X^{(01)}$, where the following definition describes the latter gate.

\begin{definition}
 The gate $X^{(ij)}$ acts on the computational basis as: \begin{equation} \label{eq:x_subspace}
    X^{(ij)}\ket{i}=\ket{j}, \; X^{(ij)}\ket{j}=\ket{i}, \; X^{(ij)}\ket{k}=\ket{k},  \text{ for } k \not \in \{i,j\}.
  \end{equation}
 
\end{definition}
So far, we introduced Clifford operators only. To generate the whole SU$(3)$, we need to introduce the following gate.

\begin{definition} \label{def:rot_s}
  The gate $R_s(\theta)$ is defined by
  \begin{equation} \label{eq:single_z_exp}
    R_s(\theta) = \exp\left(-\im \theta/2 \left(\im^s Z + \hc\right)\right),\quad \text{for}\; s \in \{0,1\},\quad 
    \theta \in \mathbb{R}.
  \end{equation}
\end{definition}
To clearly illustrate how this operator acts on the computational states, consider the following gate
\begin{equation}
  R_z^{(jk)}(\theta) = e^{-i\frac{\theta}{2}}\ket{j} \bra{j} + e^{i\frac{\theta}{2}}\ket{k} \bra{k} + \ket{m} \bra{m},\quad \text{for}\ m \notin \{j,k\}.
\end{equation}
The gate $R_s(\theta)$ decomposes into products of different $R_z^{(jk)}(\theta)$ depending on the value of $s$:
\begin{equation}
  R_0(\theta) = R_z^{(01)}(\theta) R_z^{(02)}(\theta),\quad R_1(\theta) = R_z^{(12)}(-\sqrt{3}\theta).
\end{equation}
This is relevant when single-qutrit gates have to be decomposed into a sequence of hardware native operations that act between two levels~\cite{drozhzhin2025transition}.

\begin{definition} \label{def:cx}
  The \cx gate is defined as
  \begin{equation}
      \cx \ket{i} \ket{j} = \cx \ket{i} \ket{(j+i) \bmod 3},
  \end{equation}
  and applying it $n$ times evolves two qutrits as in the following circuit:
  \begin{equation} \label{eq:entgates}
    \begin{quantikz}
      \lstick{\ket{i}} & \ctrl{1} & \rstick{\ket{i}} \\
      \lstick{\ket{j}} & \cxg{n} & \rstick{\ket{(j+ni) \bmod 3}}
    \end{quantikz}
  \end{equation}
\end{definition}
We refer to the \cx gate applied to control qutrit $i$ and target qutrit $j$ as $\cx_{ij}$, and to a generic single-qutrit gate $O$ acting on the $\ell$-th qutrit as $O_{\ell}$. The \cx gate satisfies the identities $\cx^{3} = \mathbb{1}$ and $\cx^{2} = \cx^{\dagger}$, and  the  relations
\begin{equation} \label{eq:cx_relations}
    \cx^{2m}_{12} X^n_1 \cx^{m}_{12} = X_1^{n} X_2^{nm},\quad \text{and}\quad
    \cx_{21} = (H^3 \otimes H) \cx_{12} (H \otimes H^3),
\end{equation}
for some integers $m$ and $n$. Using these relations, one can construct the qutrit \swp gate.

\begin{definition}
  The qutrit \swp gate is defined as $\swp \ket{\psi} \ket{\phi} = \ket{\phi} \ket{\psi}$. The corresponding circuit is \cite{wilmott2011, Garcia_Escartin_2013}
  \begin{equation} \label{eq:swap}
    \swp = \cx_{12} \cx^2_{21} \cx_{12} X_1^{(12)}.
  \end{equation}
\end{definition}
Note that, while the qubit \swp gate requires three \cnot gates only, the qutrit counterpart also needs an additional single-qutrit gate. The \swp gate allows distant qudits to communicate. 

The decomposition of multi-qudit unitaries into elementary gates often needs intermediary steps. For qubits, Pauli gadgets provide such an intermediary step.

\subsection{Pauli phase gadgets} \label{sec:pauli_decomp}

We refer to a Pauli string $P$ as the tensor product of $N$ qubit Pauli matrices. Any qubit Hamiltonian can be written as weighted sum of Pauli strings. We refer to the exponential $\exp(-\im P)$ as Pauli gadget, a term derived from ZX-calculus \cite{coecke2011interacting}. When $P$ corresponds to the tensor product of $N$ copies of $\sigma^z$, the Pauli gadget acts on a computational state as
\begin{equation} \label{eq:pauli_gadget}
  \exp(-\im P) \ket{\mathbf{x}} = e^{-\im \theta \left(x_1 \oplus x_2 \oplus \dots \oplus x_N \right)} \ket{\mathbf{x}},
\end{equation}
where $\bm{x}=x_1x_2\ldots x_N$ is a bit string and $\oplus$ denotes addition modulo 2. For this reason we call it a Pauli phase gadget. This unitary can be exactly decomposed into $2(N-1)$ \cnot gates and a single qubit rotation. More generally, Trotterization~\cite{nielsenchuang} approximates an arbitrary unitary as a product of Pauli gadgets. When the Pauli strings mutually commute this decomposition is exact, as is always the case for a diagonal Hamiltonian.

\section{Decomposition of qutrit diagonal unitaries and applications} \label{sec:algorithm}

\subsection{The Weyl phase gadget} \label{sec:qutrit_decomp}

  In recent years, research has focused on decomposing multi-qubit unitaries into elementary gates that are easier to implement, and on reducing the count of costly two-qubit gates. In this section we extend these results to qutrits: we decompose arbitrary qutrit diagonal unitaries into \cx gates and single-qutrit rotations. We further give the number of \cx gates required to implement the evolution of generic diagonal Hamiltonians. To this end, we propose the following generalization of Pauli phase gadgets to qutrits.

  \begin{definition} \label{def:weyl_string}
    We define the \textit{Weyl phase gadget} \wpg acting on $N$ qutrits as
    \begin{equation}
      \wpg = \exp(-\im \theta W_{s,\bm{k}}),    
    \end{equation}
    where $W_{s,\bm{k}}$ corresponds to
    \begin{equation} \label{eq:def_weylstring}
      W_{s,\bm{k}} = \im^s Z^{k_1} \otimes Z^{k_2} \otimes \dots \otimes Z^{k_N} + \hc,
      \quad s \in \{0,1\},\quad \bm{k} \in \mathcal{W},
    \end{equation}
    and $Z$ denotes the corresponding Weyl-Heisenberg qutrit operator. The set $\mathcal{W}$ is defined as
    \begin{equation} \label{eq:unique_set}
      \mathcal{W} = \{\bm{k} \in \mathbb{Z}_3^N \setminus \{\bm{0}\} : k_\ell = 1,\; \ell = \max\{j: k_j \neq 0\}\},
    \end{equation}
    where $\bm{0}=(0,0,\ldots,0)\in \mathbb{Z}_3^N$. 
    Moreover, we call \wpg a \textit{full-weight Weyl phase gadget} when $k_j \neq 0$ for all $j$.
  \end{definition}

  This constraint on $\bm{k}$ forces the last non-trivial element of the tensor product in \cref{eq:def_weylstring} to be $Z$ (and not $Z^2$). This ensure that the pair $(s,\bm{k})$ uniquely defines $W_{s,\bm{k}}$. If we remove the constraint, namely $\bm{k} \in \mathbb{Z}_3^{N}$, the pairs $(s,\bm{k})$ and $(s,2\bm{k})$ would identify the same string up to a multiplicative coefficient, since the exponent arithmetic is performed modulo 3. Restricting $\bm{k} \in \mathcal{W}$, combined with the two choices of $s \in \{0,1\}$, yields instead $3^N-1$ distinct pairs $(s,\bm{k})$, which is exactly the number of generators of the Cartan subalgebra of $\mathfrak{su}(3^N)$. The above definition of $\mathcal{W}$ is arbitrary: we could instead fix the first non-zero element of $\bm{k}$ to 1 and achieve the same result. While we focus on diagonal operators, the Weyl phase gadget can be made non-diagonal by conjugating $W_{s,\bm{k}}$ with the qutrit Hadamard using the relations after \cref{def:hadamard}.

  The qubit phase gadget in \cref{eq:pauli_gadget} applies a phase on the computational basis state proportional to the $\mathrm{XOR}$ operation on all the qubit states. While we do not pursue an analogous property here, the authors of \cite{van_de_Wetering_2023} introduced a qutrit ZX-diagram that does satisfy it, while also providing decompositions for various diagonal multi-qutrit gates. 

  Now that we have fully characterized the Weyl gadget, the next step is to decompose this operator into elementary gates, namely \cx and single-qutrit rotation gates. This decomposition is formalized in the following proposition.

  \begin{proposition}\label{lem:wey_decomp}
    A full-weight Weyl phase gadget acting on $N$ qutrits admits a circuit implementation using $2(N-1)$ \cx gates and a single-qutrit gate.
  \end{proposition}

  \begin{proof}
    Consider the relations in \cref{def:hadamard} for the qutrit Hadamard, in \cref{eq:cx_relations} for the \cx gate, and the commutation relation $ZX=\omega XZ$ introduced after Eq.~\eqref{eq:qutritXZ}. Given $j, \ell \in \{1,2\}$, the tensor product $Z^j \otimes Z^{\ell}$ can be decomposed as
    \begin{equation} \label{eq:weight_two}
      \begin{split}
        Z^{j} & \otimes Z^{\ell} = (H^3 \otimes H)(X^{j} \otimes X^{\ell})(H \otimes H^3) \\
              &= (H^3 \otimes H) \cx_{21}^{2j\ell} (\mathbb{1} \otimes X^{\ell}) \cx_{21}^{j\ell} (H \otimes H^3) \\
              &= (H^3 \otimes H) \cx_{21}^{2j\ell} (H \otimes H^3) (\mathbb{1} \otimes Z^{\ell}) (H^3 \otimes H) \cx_{21}^{j\ell} (H \otimes H^3) \\
              &= \cx_{12}^{2j\ell} (\mathbb{1} \otimes Z^{\ell}) \cx_{12}^{j\ell}.
      \end{split}
    \end{equation}
    where the second step uses the identity $\ell^2 = 1 \pmod 3$. For $N$ factors, the recursive application of the previous identity can reduce all but one of the terms in the tensor product to the identity. There are many different ways of performing this decomposition depending on the order of the recursive applications of \cref{eq:weight_two}. For simplicity, we choose the following
    \begin{equation} \label{eq:two_weyl_decomp}
      Z^{k_1} \otimes Z^{k_{2}} \otimes \dots \otimes Z^{k_N} =
      \prod_{j=1}^{N-1} \cx^{2t_i}_{j,j+1} (\id \otimes \dots \otimes \id \otimes Z^{k_N}) 
      \prod_{j=N-1}^{1} \cx^{t_i}_{j,j+1},
    \end{equation}
    for $k_i \in \{1,2\}$ and $t_i = k_i k_{i+1}$. The Hermitian conjugate of the Weyl string can be decomposed into the same sequence of \cx gates. This allows us to write the operator $W_{s,\bm{k}}$ in \cref{def:weyl_string} as
    \begin{equation} \label{eq:weyl_string_decomp}
      W_{s,\bm{k}} = \prod_{j=1}^{N-1} \cx^{2t_i} \left(i^s \id \otimes \dots \otimes \id \otimes Z + \hc\right) \prod_{j=N-1}^{1} \cx^{t_i},
    \end{equation} 
    for $t_i$ defined previously. Note that $t_{N-1} = k_{N-1}$ since, by convention, $k_{N} = 1$. The application of the property $\exp(i\theta UPU^{\dagger}) = U \exp(i \theta P) U^\dagger$, valid for any matrix $P$ and unitary $U$, to the exponential $\exp(-i \theta W_{s,\bm{k}})$, factors the CX gates out of the exponential, resulting in a circuit of $2(N-1)$ \cx gates and a single-qutrit rotation.
  \end{proof}
  The \cx count of $2(N-1)$ matches the CNOT count required for the circuit implementation of the Pauli phase gadget. The same Weyl gadget can be decomposed into different circuits, where a vector $\bm{t} \in \{1,2\}^{N-1}$ describes the exponents of the \cx gates in the entangling sequence. The vector $\bm{t}$ is in bijection with $\bm{k} \in \mathcal{W}$ whenever the index of the qutrit on which single-qutrit rotation acts is explicit.
  It follows that we can define $G(\theta, s,\bm{t}) = G(\theta, s, \bm{k})$ without ambiguity. The former notation is convenient since it makes the \cx explicit. Having $t_j = k_j k_{j+1}$, as in \cref{eq:weyl_string_decomp}, and following the convention in which operators are applied to the state from right to left, we obtain the circuit:
  \begin{equation} \label{eq:circuit_example}
    G(\theta, s, \bm{t}) =
    \scalebox{0.9}{
      \begin{quantikz}[column sep=8pt]
      & \ctrl{1} & & & & & & & & \ctrl{1} & \\
      & \cxg{t_1} & \ctrl{1} & & & & & & \ctrl{1} & \cxg{2t_1} & \\
      & & \cxg{t_2} & \ctrl{1} & & & & \ctrl{1} & \cxg{2t_2} & & \\
        \wave{} & & & & \ctrl{1} & & \ctrl{1} & & & & \\
                & & & & \cxg{t_{N-1}} & \gate{R_{s}(\theta)} & \cxg{2t_{N-1}} & & & &
    \end{quantikz}}
  \end{equation}
  where the wavy lines indicate the omitted wires. 
  
  An alternative decomposition, which can be similarly derived from \cref{eq:two_weyl_decomp}, consists in arranging all \cx gates in such a way that they target the last qutrit of the system. To give an example, consider $N=5$ qutrits and set the fifth entry of $\bm{k}$ to be 1. Then, the vector $\bm{t} \in \{0,1,2\}^{N-1}$ is now defined as $t_j = k_j$, and it identifies the exponents of the \cx gates, and the circuit associated with this decomposition is:
  \begin{equation} \label{eq:commute_cx_circuit}
    G(\theta, s, t) =
    \scalebox{0.9}{
      \begin{quantikz}[column sep=8pt]
      & \ctrl{4} & & \ \dots \ & & & & \ \dots \ & & \ctrl{4} & \\
      & & \ctrl{3} & \ \dots \ & & & & \ \dots \ & \ctrl{3} & & \\
      & & & \ \dots \ & & & & \ \dots \ & & & \\
      & & & \ \dots \ & \ctrl{1} & & \ctrl{1} & \ \dots \ & & & \\
      & \cxg{t_1} & \cxg{t_2} & \ \dots \ & \cxg{t_{N-1}} & \gate{R_{s}(\theta)} & \cxg{2t_{N-1}} & \ \dots \ & \cxg{2t_2} & \cxg{2t_1} &
    \end{quantikz}}
  \end{equation}
  This alternative decomposition allows for consecutive \cx gates to commute, since they target the same qutrit. However, it also breaks the nearest-neighbor connectivity that we instead obtain in \cref{eq:circuit_example}. Finally, we remark that equivalent circuits performing the same decomposition with a reduced circuit depth may in general exist for each specific case. Anyway, in \cref{eq:circuit_example} and \cref{eq:commute_cx_circuit} we report the most general structure of the decomposition circuits for the sake of a generic discussion.

  \cref{lem:wey_decomp} allows us to decompose the Weyl phase gadgets into elementary gates. We point out that, after the first version of this manuscript appeared, the authors of \cite{patel2025harrow} independently derived a similar protocol, which they applied to the decomposition of controlled multi-qutrit unitaries.

  \subsection{Decomposing diagonal qutrit unitaries into elementary gates}

  In the previous subsection we have introduced the qutrit Weyl phase gadget, a generalization of the Pauli phase gadget to qutrits, and provided a decomposition of this unitary into the elementary gate set composed of \cx and single-qutrit gates. We next apply these results  to decompose a generic diagonal qutrit unitary into elementary gates.
  
  First, observe that every diagonal Hamiltonian $H$ on $N$ qutrits is a real combination of operators $W_{s,\bm{k}}$, and in particular it expands as
  \begin{equation} \label{eq:ham_decomp}
    H = v_{0,0} \mathbb{1} + \sum_{s \in \{0,1\}} \sum_{\bm{k} \in \mathcal{W}} v_{s,\bm{k}}\, W_{s,\bm{k}},
  \end{equation}
  with $W_{s,\bm{k}}$ and $\mathcal{W}$ as in \cref{def:weyl_string}. The coefficients take the following values:
  \begin{equation} \label{eq:expand_coeff}
    v_{\bm{k}} = \frac{1}{3^{N}} \op{Tr}\left[\left(Z^{2k_1} \otimes Z^{2k_2} \otimes \dots \otimes Z^{2k_N}\right)H\right], \quad v_{0,\bm{k}} = \rre(v_{\bm{k}}),\quad v_{1,\bm{k}} = \rim(v_{\bm{k}}).
  \end{equation}
  This follows from the fact that, given the Weyl string $\mathcal{Z}_{\bm{k}} = Z^{k_1} \otimes \dots \otimes Z^{k_N}$ for $\bm{k} \in \mathbb{Z}_3^N$, since $\Tr(Z^a) = 3$ when $a \equiv 0 \pmod 3$ and vanishes otherwise, the operators $\{\mathcal{Z}_{\bm{k}}\}$ are Hilbert-Schmidt orthogonal:
  \begin{equation}
    \Tr(\mathcal{Z}_{\bm{j}}^\dagger \mathcal{Z}_{\bm{k}}) = \Tr(\mathcal{Z}_{\bm{k}-\bm{j}}) = 3^N \delta_{\bm{j},\bm{k}}.
  \end{equation}

  Implementing a qutrit diagonal unitary therefore reduces to implementing the Weyl phase gadgets in its decomposition. Since these all commute, they can be applied in any order, and the realization is exact. The construction is not restricted to diagonal operators. Conjugating $W_{s,\bm{k}}$ with single-qutrit gates yields non-diagonal generators, and Trotterization then approximates the evolution under a generic qutrit Hamiltonian as a product of Weyl gadgets.

  For qubits circuits, different optimization methods have been developed the reduce the number of CNOTs in a circuit \cite{welch_2014}. An upper bound on the number of gates gives a worst-case guarantee against which such optimization methods can be benchmarked. The implementation of a diagonal qubit unitary has an upper bound of $2^N - 2$ CNOTs \cite{bullockSmallerCircuitsArbitrary2003}. We derive a similar result for qutrit circuits.

  \begin{proposition} \label{pro:upper_bound}
    The implementation of a diagonal qutrit unitary acting on $N$ qutrits can be implemented using at most $(3^N - 3) / 2$ \cx gates.
  \end{proposition}
  \begin{proof}
    Consider a system of $V+1$ qutrits for $V>0$, and a matrix $t \in \{0,1,2\}^{3^V \times V}$; denote the $j$-th row of $t$ as  $t_j$. Each row $t_{j}$, together with a pair of angles $\theta_{j1},\theta_{j2} \in \mathbb{R}$, defines the circuit
    \begin{equation} \label{eq:proof_subcircuit}
      C_V=\prod_{j=1}^{3^V} G_{j,V},\quad G_{j,V} = G(\theta_{j1},0,t_j)G(\theta_{j2},1,t_j).
    \end{equation} 
    We decompose the Weyl gadgets in the definition of $G_{j,V}$ similarly to \cref{eq:commute_cx_circuit}, obtaining:
    \begin{equation} \label{eq:merge_gadget}
      G_{j,V} = \left( \prod_{i=1}^{V} \cx_{i,V+1}^{2t_{j,i}} \right) R_j 
      \left( \prod_{i=1}^{V} \cx_{i,V+1}^{t_{j,i}} \right),\;
      \text{for}\; R_j = \id_{V \times V} \otimes \left(R_0 (\theta_{j1}) R_1 (\theta_{j2})\right).
    \end{equation}
    Hence every single-qutrit rotation in $C_V$ acts on qutrit $V+1$, and the \cx gates between two rotations share this qutrit as target and therefore commute.
    Multiplying two consecutive operators gives
    \begin{equation}
      G_{j+1,V}G_{j,V} =
      \left( \prod_{i=1}^{V} \cx_{i,V+1}^{2t_{j+1,i}} \right) R_{j+1}
      \left( \prod_{i=1}^{V} \cx_{i,V+1}^{t_{j+1,i} + 2t_{j,i}} \right) R_j
      \left( \prod_{i=1}^{V} \cx_{i,V+1}^{t_{j,i}} \right)
    \end{equation}
    Since $\cx^k$ is the identity for $k = 0 \pmod 3$, the \cx count of $C_V$ is:
    \begin{equation} \label{eq:total_sum}
      T_V = \sum_{i=1}^{V} (t_{1,i}^2 \bmod 3) + \sum_{j=1}^{3^V-1} \sum_{i=1}^{V} ((2t_{j,i} 
      + t_{j+1,i})^2 \bmod 3) + \sum_{i=1}^{V} ((2t_{3^V,i})^2 \bmod 3).
    \end{equation}
    In the above expression, we have used the trick $2^2 = 1 \pmod 3$ to count each $\cx^2$ gate only once. 

    Next, to reduce this count, we choose $t_1$ to be the all-zeroes row and order the remaining rows such that $2t_j + t_{j+1}$ is nonzero in one entry only. The same applies for the last row $t_{3^V}$, which we choose to be nonzero in one entry only. This ordering is guaranteed to exist and can realized by the cyclic ternary Gray code \cite{herter201840}. Then, the first sum on the right-hand side of \cref{eq:total_sum} vanishes, the second contributes one \cx gate per row from 1 to $3^V-1$, and the third one contributes a single \cx gate:
    \begin{equation} \label{eq:proof_block_cx}
      T_V = 3^{V}.
    \end{equation}
    
    A generic diagonal unitary on $N$ qutrits factors as the product $C_1 C_2 \cdots C_{N-1}$ followed by a diagonal single-qutrit gate on the first qutrit (a single-qutrit rotation on qutrit 1 cannot be represented by $C_V$). Then, the total \cx count is given by the sum of \cref{eq:proof_block_cx} over all possible $V$:
    \begin{equation} \label{eq:total_cx}
      T = \sum_{j=1}^{N-1} T_j = \sum_{j=1}^{N-1} 3^{j} = (3^N - 3) / 2,
    \end{equation}
    which proves the proposition.
  \end{proof}

  We now provide an example of optimal decomposition. Consider the most generic $N=3$ qutrit diagonal unitary:
  \begin{equation}
    U_{\mathrm{full}} = \prod_{i=1}^{3^2 - 1} G(\theta_{i1}, 0, k_i)G(\theta_{i2}, 1, k_i),
  \end{equation}
  taken over all distinct $k_i \in \mathcal{W}$ and $\theta_{i1}, \theta_{i2} \in \mathbb{R}$. The Weyl gadgets are multiplied in the order suggested by the proof of \cref{pro:upper_bound}: the vectors $k_i$ follow the ternary Gray code order. Following this ordering and decomposing each Weyl gadget as in \cref{eq:commute_cx_circuit}, merging adjacent \cx gates yields
  \begin{equation*} \label{eq:optimal_same_target}
    U_{\mathrm{full}} =
    \scalebox{0.6}{
      \begin{quantikz}[column sep=8pt]
        & \grz{1} & \ctrl{1} & & \ctrl{1} & & \ctrl{1} & \ctrl{2} & & \ctrl{2} & & & & \ctrl{2} & & \ctrl{2} & & & & \ctrl{2} & & \ctrl{2} & & & \\
        & \grz{2} & \cxg{} & \grz{3} & \cxg{} & \grz{4} & \cxg{} & & & & & \ctrl{1} & & & & & & \ctrl{1} & & & & & & \ctrl{1} & \\
        & \grz{5} & & & & & & \cxg{} & \grz{6} & \cxg{} & \grz{7} & \cxg{} & \grz{8} & \cxg{} & \grz{9} & \cxg{} & \grz{10} & \cxg{} & \grz{11} & \cxg{} & \grz{12} & \cxg{} & \grz{13} & \cxg{} &
      \end{quantikz} 
    }
  \end{equation*}
  where we defined $R_i=R_{0}(\theta_{i1})R_{1}(\theta_{i2})$ to avoid clutter. The resulting count of 12 \cx gates agrees with \cref{pro:upper_bound}. 

We have now illustrated how to decompose a generic diagonal qutrit unitary into \cx and single-qutrit gates. In the next section we apply these methods to the qutrit-based QAOA, whose implementation requires decomposing several diagonal unitaries into elementary gates.

\subsection{Application to quantum optimization} \label{sec:decomp_application}

The quantum approximate optimization algorithm (QAOA) \cite{farhi2014} is a variational quantum algorithm designed to find approximate solutions to combinatorial problems, such as MaxCut, Traveling Salesman Problem, or graph $k$-coloring \cite{Lucas_2014, tabi2020,  Boulebnane_2024}. The algorithm begins with a uniform superposition of computational basis states over $N$ qubits, $\ket{s}=\ket{+}^{\otimes N}$. The quantum state then evolves alternately under two Hamiltonians: the cost Hamiltonian $H_c$ and the mixer Hamiltonian $H_m$. The cost Hamiltonian $H_c$ encodes the objective function of the combinatorial problem to solve, with its ground state corresponding to the optimal solution. The mixer Hamiltonian $H_m$ mixes the states, ensuring that the system does not get stuck in a local minima. It must not commute with $H_c$ to ensure effective exploration of the solution space. Different variants of QAOA employ various form of mixer Hamiltonian \cite{Hadfield_2019}, the most common choice being $H_m = \sum_{i=1}^{n} \sigma^x_i$, where $\sigma^x_i$ is the Pauli-$\sigma^x$ operator acting on the $i$-th qubit. This implementation corresponds to applying the same $x$-axis rotation gate to each qubit. The circuit alternates the two unitaries $e^{-\im \gamma_i H_c}$ and $e^{-\im \beta_i H_m}$ for $p$ layers, as depicted in Fig.~\ref{fig:qaoa_circuit}.

\begin{figure}[ht]
  \centering

  \newcommand\x{-4.45}
\newcommand\y{-0.1}
\newcommand\boxlen{7.8}
\newcommand\gx{\x+0.7}
\newcommand\gy{\y+0.6}
\newcommand\arrowh{0.9}

\definecolor{mycolor}{RGB}{255,255,255}
\definecolor{mycolor_2}{RGB}{255,255,255}
\definecolor{mycolor_3}{RGB}{0,0,0}

\tikzset{
    gateQaoa/.style={
        draw,
        inner sep=8pt,
        inner ysep=-4pt,
        minimum height=3cm
    }
}

\begin{quantikz}[wire types={q,q,n,q,n,n}, column sep=20pt]
    \lstick{$\ket{+}$} &
    \gate[4, style={gateQaoa}, disable auto height]
    {\rotatebox{90}{$\exp\left(-\im \frac{\gamma_1}{2} H_c \right)$}}
    \gategroup[4, steps=2, style={dashed, rounded corners, inner xsep=10pt, inner ysep=2pt, color=mycolor_3}]{} &
    \gate[4, style={gateQaoa}, disable auto height]
    {\rotatebox{90}{$\exp \left({-\im \frac{\beta_1}{2} H_m} \right)$}} &
    \ \dots\ &
    \gate[4, style={gateQaoa}, disable auto height]
    {\rotatebox{90}{$\exp\left(-\im \frac{\gamma_p}{2} H_c \right)$}} 
    \gategroup[4, steps=2, style={dashed, rounded corners, inner xsep=10pt, inner ysep=2pt, color=mycolor_3}]{} &
    \gate[4, style={gateQaoa}, disable auto height]
    {\rotatebox{90}{$\exp\left(-\im \frac{\beta_p}{2} H_m \right)$}} & &
    \meter{} & \setwiretype{c} \\
    \lstick{$\ket{+}$} & & & \ \dots\ & & & & \meter{} & \setwiretype{c} \\
    \lstick{$\vdots$} & & & \vdots & & & & \vdots &   \\
    \lstick{$\ket{+}$} & & & \ \dots\ & & & & \meter{} & \setwiretype{c}  \\
    & & & & & & & &  \\
    & & & & & & & \wire{c} & \cwbend{-5}
\end{quantikz}

\begin{tikzpicture}[overlay]

    \draw[->] (\gx,\gy) -- (\gx,\gy+\arrowh) node[below] {};
    \draw[->] (\gx+1.5,\gy) -- (\gx+1.5,\gy+\arrowh) node[below] {};
    \draw[->] (\gx+4.5,\gy) -- (\gx+4.5,\gy+\arrowh) node[below] {};
    \draw[->] (\gx+4.5+1.5,\gy) -- (\gx+4.5+1.5,\gy+\arrowh) node[below] {};

    \draw[rounded corners=5pt, thick] (\x,\y) rectangle (\x+\boxlen,\y+0.6);
    \node at (\x+\boxlen/2,\y+0.3) 
    {Update $(\vec{\gamma}, \vec{\beta}$) to minimize $\langle H_c \rangle$};
\end{tikzpicture}

  \caption{\label{fig:qaoa_circuit} The circuit implements the QAOA. At each iteration a classical optimizer optimizes the angles in function of the measured expectation value. The state measured from the optimized circuit provides an approximate solution to the combinatorial problem.}

\end{figure}
The unitaries depend on the angles $(\gamma_1, \dots, \gamma_p)$ and $(\beta_1, \dots, \beta_p)$, which are updated during each iteration by a classical optimizer based on the measured expectation value $\langle H_c \rangle$, aiming to minimize it. After reaching near-to-optimal parameters values, $\{\tilde{\gamma}_j, \tilde{\beta_j}\}_{j=1}^p$, the final state
\begin{equation}
  \ket{\psi(\tilde{\bm{\gamma}}, \tilde{\bm{\beta}})}=
  e^{-i \tilde{\gamma}_1 H_c} e^{-i \tilde{\beta}_1 H_m} \cdots 
  e^{-i \tilde{\gamma}_p H_c} e^{-i \tilde{\beta}_p H_m} \ket{s},
\end{equation}
is measured in the computational basis, and from the output samples one obtains approximate solutions of the problem. 

There are multiple ways to encode a combinatorial problem in the Hamiltonian $H_c$, each with its own advantages and disadvantages. In the next section we describe the one-hot and the binary encoding, highlighting how the previously discussed decomposition can be leveraged to enhance the binary encoding in a qutrit-based system.

\subsubsection{Encoding strategies} \label{sec:encoding}

The QAOA requires the combinatorial problem to be encoded within the circuit. Different encoding types can encode the same problem, each of which influence the number of qubits and the circuit depth in different way, potentially affecting the circuit's vulnerability to noise. For instance, both traveling salesman problem and the graph $\bm{k}$-coloring problem can be encoded using either the \textit{one-hot} \cite{Lucas_2014} or \textit{binary} encoding \cite{tabi2020, glos2020} schemes. We focus on the latter problem, describing its implementation in both the encoding schemes.

The graph $k$-coloring problem involves assigning a color to each node of a graph $G$ such that no adjacent nodes share the same color. The minimum number of colors required to color a graph is known as the \textit{chromatic number}. This problem has applications in network optimization, scheduling, and social networks. As an NP-complete problem \cite{garey_1990}, heuristic classical algorithms have been developed to address it \cite{brelaz_1979, LU2010241}.

The \textit{one-hot encoding} of the graph $k$-coloring problem uses $k$ qubits to encode the color of a node, as follows. Given a bit string $x$ of length $k$ that corresponds to the computational state of the system and $x_{i}$ its $i$-th component, the encoding assigns a color $v$, with $0 \leq v < k$ to the $v$-th node by setting $x_{v} = 1$ and $x_{w} = 0$ for all $w \neq v$. With the total number of qubits resulting in $N \cdot k$, with $N$ the number of nodes. For example, for $k=3$, the bit strings 100, 010, and 001 encode one of the three colors in one node, as shown in Fig.~\ref{fig:encoding:onehot}. 
The corresponding cost Hamiltonian penalizes the vertices $(v, w)$ that shares an edge and have the same color, and the bit strings $x_v$ with Hamming weight different from one.

The \textit{binary encoding} encodes the color of a node in $m=\lceil \log_2 k \rceil$ qubits. The encoding scheme takes advantage of every possible bit string combination to encode color information. For example, the graph 4-coloring uses $m=2$ qubits per node, where bit strings 00, 01, 10, 11 encode 4 different colors. Given a graph adjacency matrix $A$ the binary encoding cost Hamiltonian for the graph $k$-coloring correspond to \cite{tabi2020}:
  \begin{equation} \label{eq:binary_enc}
    H^k_C=\sum_{v,w=1}^{n} A_{vw} H^k_{v,w},\
    \quad 
    H^k_{v,w} = \sum_{\emptyset \neq S \subseteq \{0,\dots,m-1\}}
    \prod_{\ell \in S} \left( \sigma^z_{v,\ell} \sigma^z_{w,\ell} \right),
\end{equation} where $\sigma^z_{v, \ell}$ is the Pauli-Z gate applied to the $(m \cdot v + \ell)$-th qubit, and 
$a$ is a bit string of length $m$.
The Hamiltonian $H_{v,w}$ is responsible for suppressing the states where the neighboring nodes $(v, w)$ have the same color. These Hamiltonians are diagonal, therefore their exponentials commute.

When $k$ is not a power of 2, not all the bit strings represent a feasible solution for the system. The introduction of a penalty Hamiltonian $H_P$ fixes this issue by penalizing invalid bit strings. This ensures that only allowed bit strings contribute to the solution, but the implementation of $H_P$ increases the circuit depth. For example, the binary encoding for the graph 3-coloring encodes the colors in the combinations $00, 01, 10$, as showed in Fig.~\ref{fig:encoding:binary}, while $H_P = \sum_{v=1}^n\left(1-\sigma^z_{v, 1}\right)\left(1-\sigma^z_{v, 2}\right)$, penalizes the invalid state 11.

In contrast, encoding a graph ($k=d^n$)-coloring problem in a $d$-level system would only require $\lceil \log_d k \rceil$ qudits, without need of a penalty Hamiltonian.

In \cite{bottrill2023}, the authors explore a qutrit-based QAOA implementation, and compare the resources to the qubit-based QAOA, with focus on the graph 3-coloring. In the qutrit-based QAOA, the qutrit Hadamard gate in \cref{def:hadamard}, initializes each qutrit in a uniform superposition of the computational state. While the qutrit mixer Hamiltonian corresponds to the sequence $\text{R}^{(01)}_x \circ \text{R}^{(02)}_x \circ \text{R}^{(12)}_x$ applied to each qutrit.

The \textit{ternary encoding} is the natural generalization of the binary encoding to qutrit systems. Its implementation for the graph $k$-coloring encodes color of a node in 
$\lceil \log_3 k \rceil$ qutrits. Where for $k=3$ the states $\ket{0}, \ket{1}$ and $\ket{2}$ represent the colors of the nodes of the graph shown in Fig.~\ref{fig:encoding:ternary}, without requiring the penalty Hamiltonian. Additionally, in \cite{bottrill2023}, noiseless simulations demonstrate higher probabilities of sampling valid solutions compared to the binary encoding.

In the following section, we apply the methods introduced in \cref{sec:algorithm} to decompose the graph $k$-coloring cost Hamiltonian, for $k=3,9,27$, extending the theoretical results of \cite{bottrill2023}.

\begin{figure}
  \centering

  \def\ray{0.4cm}
\def\scale{1.5}
\definecolor{lightred}{RGB}{255,111,111}
\definecolor{lightgreen}{RGB}{111,255,111}
\definecolor{lightyellow}{RGB}{255,244,151}

\begin{subfigure}{.33\textwidth}
\centering
\begin{tikzpicture}
    \scriptsize

    \coordinate (circle_1) at (0,0);
    \coordinate (circle_2) at (1.5*\scale,0*\scale);
    \coordinate (circle_3) at (0.5*\scale, 0.8*\scale);
    \coordinate (circle_4) at (2*\scale, 0.8*\scale);

    \draw (circle_1) -- (circle_2);
    \draw (circle_1) -- (circle_3);
    \draw (circle_3) -- (circle_2);
    \draw (circle_3) -- (circle_4);
    \draw (circle_2) -- (circle_4);
    
    \draw[fill=lightyellow] (circle_1) circle [radius=\ray];
    \node[draw=none, inner sep=2pt] at (circle_1) {\ket{001}};
    
    \draw[fill=lightred] (circle_2) circle [radius=\ray];
    \node[draw=none, inner sep=2pt] at (circle_2) {\ket{010}};
    
    \draw[fill=lightgreen] (circle_3) circle [radius=\ray];
    \node[draw=none, inner sep=2pt] at (circle_3) {\ket{100}};
    
    \draw[fill=lightyellow] (circle_4) circle [radius=\ray];
    \node[draw=none, inner sep=2pt] at (circle_4) {\ket{001}};

\end{tikzpicture}
\caption{One-hot encoding}
\label{fig:encoding:onehot}
\end{subfigure}\begin{subfigure}{.33\textwidth}
\centering
\begin{tikzpicture}

    \coordinate (circle_1) at (0,0);
    \coordinate (circle_2) at (1.5*\scale,0*\scale);
    \coordinate (circle_3) at (0.5*\scale, 0.8*\scale);
    \coordinate (circle_4) at (2*\scale, 0.8*\scale);

    \draw (circle_1) -- (circle_2);
    \draw (circle_1) -- (circle_3);
    \draw (circle_3) -- (circle_2);
    \draw (circle_3) -- (circle_4);
    \draw (circle_2) -- (circle_4);
    
    \draw[fill=lightyellow] (circle_1) circle [radius=\ray];
    \node[draw=none, inner sep=2pt] at (circle_1) {\ket{00}};
    
    \draw[fill=lightred] (circle_2) circle [radius=\ray];
    \node[draw=none, inner sep=2pt] at (circle_2) {\ket{01}};
    
    \draw[fill=lightgreen] (circle_3) circle [radius=\ray];
    \node[draw=none, inner sep=2pt] at (circle_3) {\ket{10}};
    
    \draw[fill=lightyellow] (circle_4) circle [radius=\ray];
    \node[draw=none, inner sep=2pt] at (circle_4) {\ket{00}};

\end{tikzpicture}
\caption{Binary encoding}
\label{fig:encoding:binary}
\end{subfigure}\begin{subfigure}{0.33\textwidth}
\centering
\begin{tikzpicture}

    \coordinate (circle_1) at (0,0);
    \coordinate (circle_2) at (1.5*\scale,0*\scale);
    \coordinate (circle_3) at (0.5*\scale, 0.8*\scale);
    \coordinate (circle_4) at (2*\scale, 0.8*\scale);
    
    \draw (circle_1) -- (circle_2);
    \draw (circle_1) -- (circle_3);
    \draw (circle_3) -- (circle_2);
    \draw (circle_3) -- (circle_4);
    \draw (circle_2) -- (circle_4);
    
    \draw[fill=lightyellow] (circle_1) circle [radius=\ray];
    \node[draw=none, inner sep=2pt] at (circle_1) {\ket{0}};
    
    \draw[fill=lightred] (circle_2) circle [radius=\ray];
    \node[draw=none, inner sep=2pt] at (circle_2) {\ket{1}};
    
    \draw[fill=lightgreen] (circle_3) circle [radius=\ray];
    \node[draw=none, inner sep=2pt] at (circle_3) {\ket{2}};
    
    \draw[fill=lightyellow] (circle_4) circle [radius=\ray];
    \node[draw=none, inner sep=2pt] at (circle_4) {\ket{0}};

\end{tikzpicture}
\caption{Ternary encoding}
\label{fig:encoding:ternary}
\end{subfigure}

  \caption{\label{fig:encoding} Different encodings for the graph 3-coloring problem.
    The one-hot encoding represents a number of qubits per node equal to number of colors. In the binary encoding, $\lceil \log_2 3 \rceil=2$ qubits per node encodes the 3 colors,
    where a penalty Hamiltonian suppress the invalid state 11. Lastly, the ternary encoding uses 
  $\lceil \log_3 3 \rceil=1$ qutrits per node to perfectly encode the 3 colors.}
\end{figure}

\subsubsection{The ternary encoding}

Recent studies \cite{Deller_2023, tancara_2024} highlight the advantages of using qudit systems to solve combinatorial problems. The ternary encoding for the graph $k$-coloring problem uses $m=\lceil \log_3 k \rceil$ qutrits to represent a node. From \cref{eq:ham_decomp}, the cost Hamiltonian expands into a weighted sum of Weyl gadgets plus identity. For a generic number of colors $k$, the ternary encoding Hamiltonian  takes the form:
  \begin{equation} \label{eq:kcolorham}
    H^k_C=\sum_{v,w=1}^{n} A_{vw} H^k_{v,w},\
    \quad  H^k_{v,w} = \sum_{\emptyset \neq S \subseteq \{0,\dots,m-1\}}
    \prod_{\ell \in S} \left( Z^2_{v,\ell} Z_{w,\ell} + \hc \right),
  \end{equation}
where $Z_{v, \ell}$ represents the Weyl-Heisenberg $Z$ applied to the $(m \cdot v + \ell)$-th qutrit. We now explicitly provide the circuit decomposition of the Hamiltonian $H_{v,w}^k$ for $k=3,9,27$.

  The binary encoding for the 3-coloring problem requires 2 qubits and the implementation of a penalty Hamiltonian to suppress the fourth unwanted color, as illustrated in Fig.~4 in \cite{bottrill2023}. Instead, the ternary encoding stores the colors into the three levels of a single qutrit. For $k=3$, we derive from Eq.~\eqref{eq:kcolorham} the following Hamiltonian:
  
\begin{align} \label{eq:ham_three_color}
  H_{v,w}^{3}= Z^{2}_v Z_w + \hc.
\end{align}
  This Hamiltonian corresponds to a single Weyl gadget from \cref{def:weyl_string} whose decomposition is given by \cref{lem:wey_decomp}. The circuit implementation of its unitary evolution requires 2 \cx gates and a single-qutrit rotation:
  \begin{equation} \label{eq:3_colors}
    e^{-\im \theta H^3_{v,w}} =
    \begin{quantikz}
      \lstick{$\ket{v}$} & \ctrl{1} & & \ctrl{1} & \\
      \lstick{$\ket{w}$} & \cxg{2} & \gate{\rs} & \cxg{} &
    \end{quantikz},
  \end{equation}
for $\rs$ in \cref{def:rot_s}. The circuit is shallower than the qubit implementation for the same problem, as shown in Table~\ref{tab:resources}, and does not require any penalty Hamiltonian. 

\begin{table}[ht]
  \centering
  \begin{tabular}{r|lr|lr|cc}
    & \multicolumn{2}{c|}{Circuit depth}   & \multicolumn{2}{l|}{\# entangling gates} & \multicolumn{2}{c}{\#  qudits per node} \\ 
    \hline
    k  & Qubit  & Qutrit & Qubit    & Qutrit   & Qubit  & Qutrit  \\
    \hline
    3  & 6$\ell$+4$n$   & 3$\ell$  & 6$\ell$+4$n$   & 2$\ell$  & 2 & 1  \\
    9  & 26$\ell$+24$n$ & 8$\ell$  & 22$\ell$+28$n$ & 7$\ell$  & 4 & 2  \\
    27 & 56$\ell$+54$n$ & 29$\ell$ & 40$\ell$+60$n$ & 22$\ell$ & 5 & 3  \\
  \end{tabular}
  \caption{\label{tab:resources} The table compares different quantum resources for the implementation of the total cost Hamiltonian for the graph $k$-coloring using binary (\cref{eq:binary_enc}) and ternary (\cref{eq:kcolorham}) encodings. The coefficients $n=|V|$ and $\ell=|E|$ are, respectively, the total number of nodes and edges in the graph. The number of colors $k$ are powers of 3, which allows for an exact implementation in a qutrit system, in contrast to the qubit-based counterpart. The resources for the binary encoding are estimated by the python library t|ket> \cite{Sivarajah2020}.}
\end{table}

The binary encoding for the graph 9-coloring problem stores the color of each node using $\lceil \log_3 9 \rceil=4$ qubits. Since $2^4=16$, the penalty Hamiltonian must suppress 7 invalid colors, increasing the depth of the circuit. Instead, the problem is naturally encoded into two qutrits by ternary encoding. The qutrit Hamiltonian for the graph 9-coloring problem corresponds to
\begin{align} \label{eq:ham_9_color}
  H_{v,w}^{9} = Z^{2}_{v,0}Z_{w,0} + Z^{2}_{v,1}Z_{w,1}
  + Z_{v,0}Z^{2}_{v,1}Z^{2}_{w,0}Z_{w,1}
  + Z^{2}_{v,0}Z^{2}_{v,1}Z_{w,0}Z_{w,1} + \hc
\end{align}
Note that the first two terms also appeared in $H_{v,w}^{3}$, therefore we expect the circuit implementing these terms to be analogous to Eq.~\eqref{eq:3_colors}. The following circuit implements the unitary evolution of \cref{eq:ham_9_color}: 
  \begin{equation}
    \begin{quantikz}[column sep=10pt]
      \lstick{$\ket{v_0}$} & \ctrl{2} & & & & & & & & & \ctrl{2} & \\
      \lstick{$\ket{v_1}$} & & \ctrl{2} & & & & & & & \ctrl{2} & & \\
      \lstick{$\ket{w_0}$} & \cxg{2} & & \gate{\rs} & \ctrl{1} & & \ctrl{1} & & \ctrl{1} & & \cxg{} & \\
      \lstick{$\ket{w_1}$} & & \cxg{2} & \gate{\rs} & \cxg{2} & \gate{\rs} & \cxg{} & \gate{\rs} & \cxg{} & \cxg{} & &
    \end{quantikz}
  \end{equation}
This circuit has been obtained through the decomposition in \cref{lem:wey_decomp} followed by straightforward optimization. 

  The last problem size we consider is the graph 27-coloring. The ternary encoding for $k=27$ colors requires three qutrits per node, and its Hamiltonian contains terms including contributions from both $H_{v,w}^{3}$ and $H_{v,w}^{9}$, along with four weight-6 Weyl strings contributions:

\begin{align} \label{eq:ham_27_color}
  H_{v,w}^{27} & = \sum_{\ell=0}^2 Z^{2}_{v,\ell}Z_{w,\ell} + \sum_{p=1}^{2} \sum_{\ell=0}^{p-1} 
  (Z_{v,\ell}Z^{2}_{v,p}Z^{2}_{w,\ell}Z_{w,p}
  + Z^{2}_{v,\ell}Z^{2}_{v,p}Z_{w,\ell}Z_{w,p}) \\
               &+ \sum_{d_1, d_2 \in \{1,2\}} Z^{d_1}_{v,0}Z^{d_2}_{v,1}Z^{2}_{v,2}
               Z^{2d_1}_{w,0}Z^{2d_2}_{w,1}Z_{w,2} + \hc \nonumber 
\end{align}
The optimized circuit that implements its unitary evolution, which is again derived from \cref{lem:wey_decomp} followed by a quick optimization procedure, is illustrated in Fig.~\ref{fig:27_colors}.

\begin{figure}
  \centering
    \newcommand{\rzt}{\gate{R_0(\theta)}}
\scalebox{0.65}{
  \begin{quantikz}[column sep=5pt]
    \lstick{$\ket{v_0}$} & \ctrl{3} & & & & & & & & & & & & & & & & & \ \ldots \\
    \lstick{$\ket{v_1}$} & & \ctrl{3} & & & & & & & & & & & & & & & & \ \ldots \\
    \lstick{$\ket{v_2}$} & & & \ctrl{3} & & & & & & & & & & & & & & & \ \ldots \\
    \lstick{$\ket{w_0}$} & \cxg{2} & & & \rzt & & & & & & \ctrl{1} & & \ctrl{1} & & \ctrl{1} & \ctrl{2} & & \ctrl{2} & \ \ldots \\
    \lstick{$\ket{w_1}$} & \qw & \cxg{2} & & \rzt & \ctrl{1} & & \ctrl{1} &  & \ctrl{1} & \cxg{} & \rzt & \cxg{} & \rzt & \cxg{} & & & & \ \ldots\ \\
    \lstick{$\ket{w_2}$} & \qw & \qw & \cxg{2} & \rzt & \cxg{} & \rzt & \cxg{} & \rzt & \cxg{} & & & & & & \cxg{} & \rzt & \cxg{} & \ \ldots \\
  \end{quantikz}
}
\scalebox{0.65}{
  \begin{quantikz}[column sep=8pt]
    \ldots \ & & & & & & & & & & & & & & & & \ctrl{3} & \\
    \ldots \ & & & & & & & & & & & & & & & \ctrl{3} & & \\
    \ldots \ & & & & & & & & & & & & & & \ctrl{3} & & & \\
    \ldots \ & & \ctrl{2} & & & & & \ctrl{2} & & & & & & \ctrl{2} & & & \cxg{} & \\
    \ldots \ & & & \ctrl{1} & & \ctrl{1} & & & \ctrl{1} & & \ctrl{1} & & \ctrl{1} & & & \cxg{} & & \\
    \ldots \ & \rzt & \cxg{2} & \cxg{} & \rzt & \cxg{} & \rzt & \cxg{} & \cxg{} & \rzt & \cxg{} & \rzt & \cxg{} & \cxg{1} & \cxg{} & & & \\
  \end{quantikz}
}
  \caption{\label{fig:27_colors} The circuit implements the graph 27-coloring Hamiltonian in Eq.~\eqref{eq:ham_27_color}.}
\end{figure}

Our analysis shows that, for the considered values of $k$, the qutrit implementation achieves a notable reduction in circuit depth and gate count when compared to its qubit counterpart, as shown in \cref{tab:resources}. This reduction appears to become more pronounced with increasing $k$.

\section{Extraction of qutrit circuits in restricted topologies} \label{sec:restrictedTop}

\subsection{Restricted topologies}

In the previous chapter we introduced a decomposition scheme for qutrit diagonal unitaries with application to optimization algorithms. We assumed all-to-all communication among qutrits, where multi-qutrit gates were allowed to entangle each pair of qutrits. In this chapter we consider different topologies for which not all qutrits are connected to each other.

Common connectivity topologies include linear nearest-neighbor, square lattice, heavy-hex lattice, and star topology, where limited connectivity constrains interaction between qubits (or qudits). Communication between distant qudits is typically enabled by the application of SWAP gates.

The use of SWAP gates, however, can significantly increase circuit depth and two-qudit gate count. To reduce the effect of limited qubit connectivity on circuit depth, various methods have been developed. A prominent example is the \steinergauss algorithm ~\cite{kissinger2019}, which addresses the connectivity constraint while achieving a lower CNOT gate count than alternative methods. The algorithm applies to circuits composed solely of CNOTs. Our extension to qutrits will equivalently apply to circuits composed of CX gates only, exactly as those that appear in the first and second half of the circuit in  \cref{eq:weyl_string_decomp}.

In the following, we outline this algorithm and introduce its generalization to qutrit systems with restricted topologies. A key theoretical contribution of our work is the extension of the \textit{parity matrix} framework to qutrits. While this generalization introduces some minor modifications to the original method, its core principles transfer seamlessly.

\subsection{The Steiner tree and the parity map} \label{sec:steiner_review}

The \textit{Steiner tree} problem takes as input a subset of vertices of a graph $G$, called \textit{terminal vertices}, and seeks to find a tree $T$ within $G$ that connects all the \textit{terminal vertices}. The goal is to minimize the number of vertices in $T$. This problem is known to be NP-complete, and heuristic algorithms are commonly used to find good approximations of the optimal solution \cite{Robins2000}. Any vertex $v \in T$ that is not a terminal vertex is referred to as a \textit{Steiner vertex}. 

We also consider a variant of this problem. The \textit{decreasing Steiner tree} problem takes as input an undirected graph $G$ whose vertices are totally ordered under a binary relation $\leq$, and a subset of vertices called \textit{terminal vertices}; then, it seeks to find a \textit{decreasing Steiner tree} $T$, which is a rooted tree within $G$ that satisfies the following condition: for every vertex $v \in T$, all of its children in the tree are strictly less than $v$ under the ordering $\leq$. A vertex $v_a$ is considered a child of another vertex $v_b$ if they are adjacent and the distance (in number of edges) between $v_a$ to the root of $T$ is exactly one more than the distance from $v_b$ to the root. The \textit{decreasing Steiner tree} can only be applied to graphs that contain a Hamiltonian path, i.e., a path that visits each vertex exactly once. We consider only graphs satisfying this property, while methods that extend the algorithm to general graphs exist \cite{kissinger2019}.

A parity map is an invertible linear map on $N$-bit strings, represented by an invertible $N \times N$ matrix over the finite field $\mathbb{Z}_2$. Parity maps provide an intermediate representation for circuits built entirely from CNOT gates. Each parity map admits a synthesis into a sequence of CNOT operations, although this synthesis is not unique.

Given a parity matrix $P$ and assuming all-to-all connectivity, the following steps extract the CNOT sequence from the map $P$:
\begin{itemize}
  \item Define $P_j$ as the $j$-th row of $P$, with coefficients $P_{jk}$. Also, initialize a quantum circuit $C$ that acts as the identity on $N$ qubits.
  \item For each column $k$, starting from $k=1$ to $N$, perform the following: For every $j > k$ such that $P_{jk}=1$, perform the row operation $P_j := P_j + P_k$ and append the CNOT$_{j.k}$ gate in the circuit $C$. $P$ is now a upper triangular matrix.
  \item For each column $d$ from $N$ to 1, perform: For every $j < k$ such that $P_{jk}=1$, perform the row operation $P_j := P_j + P_k$ and append the CNOT$_{j,k}$ gate in the circuit $C$. This eliminates the upper-triangular entries and reduces $P$ to the identity matrix.
\end{itemize}
The quantum circuit $C^{-1}$ is the extracted circuit, and implements the same bit-string to bit-string mapping as the original parity map $P$. Note that the extracted circuit is not unique, different orderings of the row operations can lead to different circuit with different depths and gate counts. An optimized ordering can significantly reduce the total number of CNOT gates \cite{ketan2004}. An examples of a parity matrix and corresponding CNOT-only circuit is the following:

\begin{equation*}
  P=\left(\begin{matrix}
      1 & 1 & 1 & 0 \\
      0 & 1 & 1 & 0 \\
      1 & 0 & 1 & 1 \\
      1 & 0 & 0 & 1
  \end{matrix}\right)
  \qquad \qquad
  \begin{quantikz}
    \lstick{$\ket{\psi_0}$} & \ctrl{1} & & \targ{} & 
    \rstick{$\ket{\psi_3 \oplus \psi_2 \oplus \psi_0}$}\\
    \lstick{$\ket{\psi_1}$} & \targ{} & \ctrl{1} & &
    \rstick{$\ket{\psi_1 \oplus \psi_0}$} \\
    \lstick{$\ket{\psi_2}$} & \ctrl{1} & \targ{} & &
    \rstick{$\ket{\psi_2 \oplus \psi_1 \oplus \psi_0}$} \\
    \lstick{$\ket{\psi_3}$} & \targ{} & & \ctrl{-3} &
    \rstick{$\ket{\psi_3 \oplus \psi_2}$}
  \end{quantikz}
\end{equation*}

For restricted topologies the circuit extraction requires additional steps. The \steinergauss algorithm addresses this by constraining row operations respect to the connectivity of the underlying architecture. 
It consists of two subroutines, \steinerdown and \steinerup. The \steinerdown subroutine uses a Steiner tree to restrict the allowed row operations, while \steinerup applies the same idea on the decreasing Steiner tree. Before describing this algorithm, we introduce the ternary parity map, a generalization of the parity map to qutrit systems.

\subsection{The ternary parity map as intermediate representation} \label{sec:ternary_routing}

We define the \textit{ternary parity map} $P$ as a $N \times N$ matrix over $\mathbb{Z}_3$ that provides an invertible linear map from ternary strings to ternary strings, and acts as an intermediate representation for qutrit circuits that contain exclusively \cx and $X^{12}$ gates. Note, the latter gate is equivalent to a sequence of \cx gates through an ancilla qutrit.

The circuit extraction process for an all-to-all topology closely resembles the CNOT-only circuit extraction in the qubit case, but with important distinctions arising from ternary logic. Given a ternary parity map $P$ and row indexes $i,j$ with $i \neq j$, when performing the row operation $P_j := P_j + P_i$, the gate $\cx_{i,j}$ is applied to the circuit; instead, for the row operation $P_j := P_j - P_i$\red{,} the gate $\cx^2_{i,j}$ is applied. Moreover, the row operation $P_j := 2P_j$, that exchanges the matrix element 1 with 2, corresponds to the gate $X^{(12)}$ acting on the $j$-th qutrit. An example of a ternary parity map and respective circuit is:
\begin{equation}
  P=\left(\;\begin{matrix}
      1 & 2 & 0 & 2 \\
      0 & 1 & 0 & 0 \\
      1 & 2 & 2 & 1 \\
      2 & 0 & 0 & 1
  \end{matrix}\;\right)
  \rightarrow
  \scalebox{0.8}{
    \begin{quantikz}[column sep=6pt]
      \lstick{$\ket{\psi_1}$} & \cxg{2} & \cxg{} & & & & & \ctrl{3} & \ctrl{1} &
      \rstick{$\ket{2\psi_4 + \psi_3 + \psi_1 }$}\\
      \lstick{$\ket{\psi_2}$} & \ctrl{-1} & & & \ctrl{2} & \ctrl{1} & \cxg{} & & \cxg{2} &
      \rstick{$\ket{2\psi_3 + \psi_2 + 2\psi_1}$} \\
      \lstick{$\ket{\psi_3}$} & & \ctrl{-2} & \cxg{2} & & \cxg{2} & \ctrl{-1} & & 
      \xonetwo & \rstick{$\ket{2\psi_3}$} \\
      \lstick{$\ket{\psi_4}$} & & & \ctrl{-1} & \cxg{2} & & & \cxg{2} & &
      \rstick{$\ket{\psi_4 + \psi_3 + 2\psi_1}$}
  \end{quantikz}}
\end{equation}
We extract the circuit by performing any sequence of row operations that makes $P$ equal to the identity. Given the row indexes $i$ and $j$ for $i \neq j$, the row operations correspond to the gates:
\begin{equation} \label{eq:row_to_gates}
  P_j := P_j + P_i\; \rightarrow\; \cx_{i,j},\quad P_j := P_j - P_i\; \rightarrow\; \cx^2_{i,j},\quad P_j := 2P_j\; \rightarrow\; X^{(12)}_j.
\end{equation}
After the application of a row operation, we append the associated gate to the beginning of the initially empty circuit, after the last row operation, that makes $P=\id$, the circuit is complete.

We now extend the \steinergauss algorithm to qutrit systems. Let $P$ be a ternary parity map and $G$ a graph of order $N$ describing the topology, whose vertices are labeled from 1 to $N$. The following \steinergauss subroutines apply a sequence of row operations that makes $P$ diagonal while respecting the given constraints:
\begin{enumerate}
  \item
    \steinerdown: Let $P_{j}$ denote the $j$-th row of $P$, with coefficients $P_{jk}$ and $j \geq k$. Processing the columns $k$ in increasing order, take as terminals the nodes $j$ for $P_{jk} \neq 0$ and the node $k$; then, compute the corresponding Steiner tree and label the computed Steiner tree vertices as $i_1, \ldots, i_n$. If the $P_{kk}$ entry is zero, first restore it by adding an adjacent row, $P_{i_1} := P_{i_1} + P_{i_2}$, so that $P_{kk} \neq 0$. Then perform the row operations:
    \begin{equation} \label{eq:steiner_down}
      P_{i_w} := P_{i_w} + \sum_{t=1}^{w-1} \alpha_t P_{i_{t}},\quad \text{for}\; \bm{\alpha} \in \{1,2\}^{w-1}\quad \text{and}\; 1 < w \leq n,
    \end{equation}
    where, for each $w$, the coefficients $\alpha_t \in \mathbb{Z}_3$ are the choice such that $P_{i_w k} = 0$ after the update. The operations are applied in a decreasing order of $w$. After repeating this for every column $k$, the ternary parity map $P$ is in upper-triangular form and ready for the next step.

  \item 
    \steinerup: Start from the last column, compute the decreasing Steiner tree in $G$ for terminal vertices $j$ such that $P_{jN} \neq 0$ with $j<N$, and vertex $N$. This variant of the Steiner tree preserves the upper-triangular form of $P$, since the child of each node is labeled with a lower index. As in the previous step, label the Steiner vertices as $i_1, \dots, i_n$ and perform the same row operations already described in the former equation. Repeat the process for every $j$. Every coefficient of $P_N$ is now 0 expect for $P_{NN}$. Repeat the process for every column in decreasing order. The map $P$ is now diagonal.

  \item For each diagonal element $P_{jj} = 2$, perform the row operation $P_j := 2P_j$. As a result, $P = \mathbb{1}$.

\end{enumerate}
To extract the circuit $C$ we append to the initially empty  circuit the gates associated with each row operation. The circuit inverse $C^{-1}$ implements the same mapping of the original ternary parity map $P$. As an example, in the next section we describe step-by-step the circuit extraction procedure to a topology reassembling a grid.

\subsection{Example}

The example provided here is similar to the one that can be found in \cite{kissinger2019}. Let us consider the following topology graph $G$ and ternary parity map $P$:
\begin{center}
  \begin{minipage}{.4\linewidth}
    \begin{tikzpicture}

      \node[draw=none, inner sep=2pt] at (-1, 1) {$G \equiv$};

      \draw[line width=0.03cm] (0,0) -- (1,0);
      \draw[line width=0.03cm] (1,0) -- (2,0);
      \draw[line width=0.03cm] (0,1) -- (1,1);
      \draw[line width=0.03cm] (1,1) -- (2,1);
      \draw[line width=0.03cm] (0,2) -- (1,2);
      \draw[line width=0.03cm] (1,2) -- (2,2);

      \draw[line width=0.03cm] (0,0) -- (0,1);
      \draw[line width=0.03cm] (0,1) -- (0,2);
      \draw[line width=0.03cm] (1,0) -- (1,1);
      \draw[line width=0.03cm] (1,1) -- (1,2);
      \draw[line width=0.03cm] (2,0) -- (2,1);
      \draw[line width=0.03cm] (2,1) -- (2,2);

      \foreach \i in {0, 1, 2} {
        \foreach \j in {0, 1, 2} {

          \pgfmathtruncatemacro{\label}{abs((\j-2))*3 + \i + 1}

          \ifthenelse{\j=1}
          {\pgfmathtruncatemacro{\label}{abs((\j-2))*3 + abs(\i-2) +1}}{}

          \filldraw[fill=black] (\j, \i) circle (0.11cm);
          \node[draw=none, inner sep=2pt] at (\i-0.30, \j+0.25) {\label};

        }

      }
    \end{tikzpicture}
  \end{minipage}
  \begin{minipage}{.43\linewidth}
    \begin{equation*}
      P= \scalemath{0.75}{\left(\;\begin{matrix}
            1 & 0 & 0 & 1 & 0 & 0 & 0 & 1 & 0 \\
            0 & 1 & 0 & 2 & 0 & 1 & 1 & 0 & 0 \\
            2 & 0 & 1 & 0 & 0 & 1 & 1 & 0 & 1 \\
            0 & 1 & 0 & 1 & 0 & 2 & 0 & 0 & 0 \\
            0 & 0 & 0 & 0 & 2 & 0 & 0 & 0 & 2 \\
            0 & 0 & 0 & 0 & 0 & 2 & 0 & 1 & 0 \\
            0 & 0 & 0 & 0 & 2 & 0 & 1 & 0 & 0 \\
            2 & 1 & 1 & 2 & 2 & 0 & 0 & 1 & 0 \\
            0 & 0 & 2 & 2 & 2 & 0 & 0 & 0 & 1
      \end{matrix}\; \right)}
    \end{equation*}
  \end{minipage}
\end{center}

Note that $G$ contains an Hamiltonian path that cross each vertex. We apply the algorithm illustrated in the previous section to extract the respective circuit $C$ from $P$. Following the \steinerdown step, we compute the Steiner tree of the first column of $P$ with Steiner terminals $S=\{1, 3, 8\}$, and choose vertex 1 as the root of the tree. The corresponding Steiner tree $T$ is shown on the right:
\definecolor{gray}{RGB}{150,150,150}
\def\linew{0.04cm}
\def\circlerad{0.11cm}
\def\gridspace{0.03cm}

\begin{center}
\begin{minipage}{0.3\textwidth}
\raggedright
\begin{tikzpicture}
    
    \draw[line width=\gridspace] (0,0) -- (1,0);
    \draw[line width=\gridspace] (1,0) -- (2,0);
    \draw[line width=\gridspace] (0,1) -- (1,1);
    \draw[line width=\gridspace] (1,1) -- (2,1);
    \draw[line width=\gridspace] (0,2) -- (1,2);
    \draw[line width=\gridspace] (1,2) -- (2,2);

    \draw[line width=\gridspace] (0,0) -- (0,1);
    \draw[line width=\gridspace] (0,1) -- (0,2);
    \draw[line width=\gridspace] (1,0) -- (1,1);
    \draw[line width=\gridspace] (1,1) -- (1,2);
    \draw[line width=\gridspace] (2,0) -- (2,1);
    \draw[line width=\gridspace] (2,1) -- (2,2);
    
        

            



    \foreach \i in {0, 1, 2} {
        \foreach \j in {0, 1, 2} {

            \pgfmathtruncatemacro{\label}{abs((\j-2))*3 + \i + 1}

            \ifthenelse{\j=1}
                {\pgfmathtruncatemacro{\label}{abs((\j-2))*3 + abs(\i-2) + 1}}{}
            
            \filldraw[fill=black] (\j, \i) circle (\circlerad);
            \node[draw=none, inner sep=2pt] at (\i-0.33, \j+0.27) {\label};

            \ifthenelse{\i=2 \AND \j=0}{
                \draw[draw=red, line width=\linew] (\j, \i) circle [radius=0.2cm];
                \filldraw[fill=black] (\j, \i) circle (\circlerad);
            }{}

            \ifthenelse{\(\i=2 \AND \j=2\) \OR \(\i=0 \AND \j=1\)}{
                \draw[draw=blue, line width=\linew] (\j, \i) circle [radius=0.2cm];
                \filldraw[fill=black] (\j, \i) circle (\circlerad);
            }{}
        }
    }
\end{tikzpicture}
\end{minipage}\begin{minipage}{0.1\textwidth}
\centering
\begin{tikzpicture}
\draw[->] (0,0)   -- (1,0);
\end{tikzpicture}
\end{minipage}\begin{minipage}{0.3\textwidth}
\centering
\begin{tikzpicture}
    \node[draw=none, inner sep=2pt] at (-1, 1) {$T \equiv$};

    \draw[gray, line width=\gridspace] (0,0) -- (1,0);
    \draw[gray, line width=\gridspace] (1,0) -- (2,0);
    \draw[gray, line width=\gridspace] (0,1) -- (1,1);
    \draw[gray, line width=\gridspace] (1,1) -- (2,1);
    \draw[line width=\gridspace] (0,2) -- (1,2);
    \draw[line width=\gridspace] (1,2) -- (2,2);

    \draw[gray, line width=\gridspace] (0,0) -- (0,1);
    \draw[gray, line width=\gridspace] (0,1) -- (0,2);
    \draw[line width=\gridspace] (1,0) -- (1,1);
    \draw[line width=\gridspace] (1,1) -- (1,2);
    \draw[gray, line width=\gridspace] (2,0) -- (2,1);
    \draw[gray, line width=\gridspace] (2,1) -- (2,2);
    
    \foreach \i in {0, 1, 2} {
        \foreach \j in {0, 1, 2} {
        
            \pgfmathtruncatemacro{\label}{abs((\j-2))*3 + \i + 1}

            \ifthenelse{\j=1}
                {\pgfmathtruncatemacro{\label}{abs((\j-2))*3 + abs(\i-2) + 1}}{}
            
            \filldraw[fill=gray] (\j, \i) circle (\circlerad);
            \node[draw=none, inner sep=2pt] at (\i-0.33, \j+0.27) {\label};

            \ifthenelse{\i=2 \AND \j=0}{
                \draw[draw=red, line width=\linew] (\j, \i) circle [radius=0.2cm];
                \filldraw[fill=black] (\j, \i) circle (\circlerad);
            }{}

            \ifthenelse{\(\i=2 \AND \j=2\) \OR \(\i=0 \AND \j=1\)}{
                \draw[draw=blue, line width=\linew] (\j, \i) circle [radius=0.2cm];
                \filldraw[fill=black] (\j, \i) circle (\circlerad);
            }{}

            \ifthenelse{\(\i=1 \AND \j=1\) \OR \(\i=2 \AND \j=1\)}{
                \draw[draw=brown, line width=\linew] (\j, \i) circle [radius=0.2cm];
                \filldraw[fill=black] (\j, \i) circle (\circlerad);
            }{}

        }
        
    }
\end{tikzpicture}
\end{minipage}
\end{center}

The terminal vertices are both those circled in blue and the root vertex circled in red. The brown vertices are Steiner points, which mediate interactions between terminal nodes. This tree structure yields the minimum number of row operations on $P$ required to set to zero the coefficients $P_{31}$ and $P_{81}$, without altering the other entries in the column. The following sequence of operations achieves this:
\newcommand{\myscale}{0.55}
\begin{align*}
  P=& 
  \scalemath{\myscale}{
    \left(\;\begin{matrix}
        1 & 0 & 0 & 1 & 0 & 0 & 0 & 1 & 0 \\
        0 & 1 & 0 & 2 & 0 & 1 & 1 & 0 & 0 \\
        2 & 0 & 1 & 0 & 0 & 1 & 1 & 0 & 1 \\
        0 & 1 & 0 & 1 & 0 & 2 & 0 & 0 & 0 \\
        0 & 0 & 0 & 0 & 2 & 0 & 0 & 0 & 2 \\
        0 & 0 & 0 & 0 & 0 & 2 & 0 & 1 & 0 \\
        0 & 0 & 0 & 0 & 2 & 0 & 1 & 0 & 0 \\
        2 & 1 & 2 & 2 & 2 & 0 & 0 & 1 & 0 \\
        0 & 0 & 2 & 2 & 2 & 0 & 0 & 0 & 1
  \end{matrix}\; \right)}
  \xrightarrow[\text{CX}_{1,2}]{P_2:=P_2+P_1}
  \scalemath{\myscale}{
    \left(\;\begin{matrix}
        1 & 0 & 0 & 1 & 0 & 0 & 0 & 1 & 0 \\
        1 & 1 & 0 & 0 & 0 & 1 & 1 & 1 & 0 \\
        2 & 0 & 1 & 0 & 0 & 1 & 1 & 0 & 1 \\
        0 & 1 & 0 & 1 & 0 & 2 & 0 & 0 & 0 \\
        0 & 0 & 0 & 0 & 2 & 0 & 0 & 0 & 2 \\
        0 & 0 & 0 & 0 & 0 & 2 & 0 & 1 & 0 \\
        0 & 0 & 0 & 0 & 2 & 0 & 1 & 0 & 0 \\
        2 & 1 & 2 & 2 & 2 & 0 & 0 & 1 & 0 \\
        0 & 0 & 2 & 2 & 2 & 0 & 0 & 0 & 1
  \end{matrix}\; \right) }
  \xrightarrow[\text{CX}_{2,3}]{P_3:=P_3+P_2}
  \scalemath{\myscale}{\left(\;\begin{matrix}
        1 & 0 & 0 & 1 & 0 & 0 & 0 & 1 & 0 \\
        1 & 1 & 0 & 0 & 0 & 1 & 1 & 1 & 0 \\
        0 & 1 & 1 & 0 & 0 & 2 & 2 & 1 & 1 \\
        0 & 1 & 0 & 1 & 0 & 2 & 0 & 0 & 0 \\
        0 & 0 & 0 & 0 & 2 & 0 & 0 & 0 & 2 \\
        0 & 0 & 0 & 0 & 0 & 2 & 0 & 1 & 0 \\
        0 & 0 & 0 & 0 & 2 & 0 & 1 & 0 & 0 \\
        2 & 1 & 2 & 2 & 2 & 0 & 0 & 1 & 0 \\
        0 & 0 & 2 & 2 & 2 & 0 & 0 & 0 & 1
  \end{matrix}\; \right)} \notag \\
  \xrightarrow[\cx^{2}_{2,5}]{P_5:=P_5-P_2}
    &
    \scalemath{\myscale}{
      \begin{pmatrix}
        1 & 0 & 0 & 1 & 0 & 0 & 0 & 1 & 0 \\
        1 & 1 & 0 & 0 & 0 & 1 & 1 & 1 & 0 \\
        0 & 1 & 1 & 0 & 0 & 2 & 2 & 1 & 1 \\
        0 & 1 & 0 & 1 & 0 & 2 & 0 & 0 & 0 \\
        2 & 2 & 0 & 0 & 2 & 2 & 2 & 2 & 2 \\
        0 & 0 & 0 & 0 & 0 & 2 & 0 & 1 & 0 \\
        0 & 0 & 0 & 0 & 2 & 0 & 1 & 0 & 0 \\
        2 & 1 & 2 & 2 & 2 & 0 & 0 & 1 & 0 \\
        0 & 0 & 2 & 2 & 2 & 0 & 0 & 0 & 1
    \end{pmatrix}}\notag
    \xrightarrow[\cx^2_{5,8}]{P_8:=P_8-P_5}
    \scalemath{\myscale}{
      \left(\;\begin{matrix}
          1 & 0 & 0 & 1 & 0 & 0 & 0 & 1 & 0 \\
          1 & 1 & 0 & 0 & 0 & 1 & 1 & 1 & 0 \\
          0 & 1 & 1 & 0 & 0 & 2 & 2 & 1 & 1 \\
          0 & 1 & 0 & 1 & 0 & 2 & 0 & 0 & 0 \\
          2 & 2 & 0 & 0 & 2 & 2 & 2 & 2 & 2 \\
          0 & 0 & 0 & 0 & 0 & 2 & 0 & 1 & 0 \\
          0 & 0 & 0 & 0 & 2 & 0 & 1 & 0 & 0 \\
          0 & 2 & 2 & 2 & 0 & 1 & 1 & 2 & 2 \\
          0 & 0 & 2 & 2 & 2 & 0 & 0 & 0 & 1
    \end{matrix}\; \right)}
    \xrightarrow[\cx_{2,5}]{P_5:=P_5+P_2}
    \scalemath{\myscale}{
      \left(\;\begin{matrix}
          1 & 0 & 0 & 1 & 0 & 0 & 0 & 1 & 0 \\
          1 & 1 & 0 & 0 & 0 & 1 & 1 & 1 & 0 \\
          0 & 1 & 1 & 0 & 0 & 2 & 2 & 1 & 1 \\
          0 & 1 & 0 & 1 & 0 & 2 & 0 & 0 & 0 \\
          0 & 0 & 0 & 0 & 2 & 1 & 1 & 1 & 2 \\
          0 & 0 & 0 & 0 & 0 & 2 & 0 & 1 & 0 \\
          0 & 0 & 0 & 0 & 2 & 0 & 1 & 0 & 0 \\
          0 & 2 & 2 & 2 & 0 & 1 & 1 & 2 & 2 \\
          0 & 0 & 2 & 2 & 2 & 0 & 0 & 0 & 1
    \end{matrix}\; \right)}\\
  \xrightarrow[\cx^{2}_{1,2}]{P_2:=P_2-P_1} & 
  \scalemath{\myscale}{
    \left(\;\begin{matrix}
        1 & 0 & 0 & 1 & 0 & 0 & 0 & 1 & 0 \\
        0 & 1 & 0 & 2 & 0 & 1 & 1 & 0 & 0 \\
        0 & 1 & 1 & 0 & 0 & 2 & 2 & 1 & 1 \\
        0 & 1 & 0 & 1 & 0 & 2 & 0 & 0 & 0 \\
        0 & 0 & 0 & 0 & 2 & 1 & 1 & 1 & 2 \\
        0 & 0 & 0 & 0 & 0 & 2 & 0 & 1 & 0 \\
        0 & 0 & 0 & 0 & 2 & 0 & 1 & 0 & 0 \\
        0 & 2 & 2 & 2 & 0 & 1 & 1 & 2 & 2 \\
        0 & 0 & 2 & 2 & 2 & 0 & 0 & 0 & 1
  \end{matrix}\; \right)}
  \xrightarrow{\hspace{0.3cm}} \cdots \xrightarrow{\hspace{0.3cm}}
  \scalemath{\myscale}{
    \left(\;\begin{matrix}
        1 & 0 & 0 & 1 & 0 & 0 & 0 & 1 & 0 \\
        0 & 2 & 0 & 0 & 0 & 1 & 1 & 0 & 0 \\
        0 & 0 & 1 & 0 & 0 & 2 & 2 & 1 & 1 \\
        0 & 0 & 0 & 1 & 0 & 2 & 0 & 0 & 0 \\
        0 & 0 & 0 & 0 & 2 & 1 & 1 & 1 & 2 \\
        0 & 0 & 0 & 0 & 0 & 1 & 0 & 1 & 0 \\
        0 & 0 & 0 & 0 & 0 & 0 & 1 & 0 & 0 \\
        0 & 0 & 0 & 0 & 0 & 0 & 0 & 1 & 0 \\
        0 & 0 & 0 & 0 & 0 & 0 & 0 & 0 & 1
  \end{matrix}\; \right)}
\end{align*}

The gate related to each row operation is shown below the corresponding arrow. After clearing the first column we obtain the circuit:
\begin{equation}
  C = \cx_{1,2} \cx_{2,3} \cx_{2,5}^{2} \cx_{5,8}^2 \cx_{2,5} \cx_{1,2}^2.
\end{equation}
We conclude the first step by computing a Steiner tree for the remaining columns $1 < i \leq 9$ in increasing order, with the terminal vertices $j$ such that $P_{ji} = 0$ for $i \leq j$. After performing each row operation, $P$ becomes an upper triangular.

The \steinerup step transforms the upper triangular matrix $P$ into a diagonal matrix through row operations that satisfy the given topology. We calculate the decreasing Steiner tree problem for the last column of $P$. The decreasing tree for the last column has terminal nodes $S=\{9, 5, 3\}$ and corresponds to:
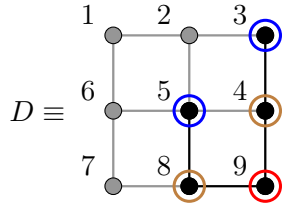
\definecolor{gray}{RGB}{150,150,150}
\begin{center}
\def\linew{0.04cm}

\begin{tikzpicture}
    \node[draw=none, inner sep=2pt] at (-1, 1) {$D \equiv$};

    \draw[gray, line width=0.03cm] (0,0) -- (1,0); 
    \draw[line width=0.03cm] (1,0) -- (2,0); 
    \draw[gray, line width=0.03cm] (0,1) -- (1,1); 
    \draw[gray, line width=0.03cm] (1,1) -- (2,1); 
    \draw[gray, line width=0.03cm] (0,2) -- (1,2); 
    \draw[gray, line width=0.03cm] (1,2) -- (2,2); 

    \draw[gray, line width=0.03cm] (0,0) -- (0,1); 
    \draw[gray, line width=0.03cm] (0,1) -- (0,2); 
    \draw[line width=0.03cm] (1,0) -- (1,1); 
    \draw[gray, line width=0.03cm] (1,1) -- (1,2); 
    \draw[line width=0.03cm] (2,0) -- (2,1); 
    \draw[line width=0.03cm] (2,1) -- (2,2); 
    
    \foreach \i in {0, 1, 2} {
        \foreach \j in {0, 1, 2} {
        
            \pgfmathtruncatemacro{\label}{abs((\j-2))*3 + \i +1}

            \ifthenelse{\j=1}
                {\pgfmathtruncatemacro{\label}{abs((\j-2))*3 + abs(\i-2) +1}}{}
            
            \filldraw[fill=gray] (\j, \i) circle (0.11cm);
            \node[draw=none, inner sep=2pt] at (\i-0.33, \j+0.27) {\label};

            \ifthenelse{\i=0 \AND \j=2}{
                \draw[draw=red, line width=\linew] (\j, \i) circle [radius=0.2cm];
                \filldraw[fill=black] (\j, \i) circle (0.11cm);
            }{}

            \ifthenelse{\(\i=2 \AND \j=2\) \OR \(\i=1 \AND \j=1\)}{
                \draw[draw=blue, line width=\linew] (\j, \i) circle [radius=0.2cm];
                \filldraw[fill=black] (\j, \i) circle (0.11cm);
            }{}

            \ifthenelse{\(\i=1 \AND \j=2\) \OR \(\i=0 \AND \j=1\)}{
                \draw[draw=brown, line width=\linew] (\j, \i) circle [radius=0.2cm];
                \filldraw[fill=black] (\j, \i) circle (0.11cm);
            }{}
        }
    }
\end{tikzpicture}
\end{center}
As in the previous step, we perform the minimum number of row operations that set the coefficients $P_{93}=0$ and $P_{95}=0$. These operations must also preserve the upper triangular form, a property guaranteed by the decreasing Steiner tree. This is achieved by the following row operations:
\begin{align*}
  P=& 
  \scalemath{\myscale}{
    \left(\;\begin{matrix}
        1 & 0 & 0 & 1 & 0 & 0 & 0 & 1 & 0 \\
        0 & 2 & 0 & 0 & 0 & 1 & 1 & 0 & 0 \\
        0 & 0 & 1 & 0 & 0 & 2 & 2 & 1 & 1 \\
        0 & 0 & 0 & 1 & 0 & 2 & 0 & 0 & 0 \\
        0 & 0 & 0 & 0 & 2 & 1 & 1 & 1 & 2 \\
        0 & 0 & 0 & 0 & 0 & 1 & 0 & 1 & 0 \\
        0 & 0 & 0 & 0 & 0 & 0 & 1 & 0 & 0 \\
        0 & 0 & 0 & 0 & 0 & 0 & 0 & 1 & 0 \\
        0 & 0 & 0 & 0 & 0 & 0 & 0 & 0 & 1
  \end{matrix}\; \right)}
  \xrightarrow[\cx_{9,4}]{P_4:=P_4+P_9}
  \scalemath{\myscale}{
    \left(\;\begin{matrix}
        1 & 0 & 0 & 1 & 0 & 0 & 0 & 1 & 0 \\
        0 & 2 & 0 & 0 & 0 & 1 & 1 & 0 & 0 \\
        0 & 0 & 1 & 0 & 0 & 2 & 2 & 1 & 1 \\
        0 & 0 & 0 & 1 & 0 & 2 & 0 & 0 & 1 \\
        0 & 0 & 0 & 0 & 2 & 1 & 1 & 1 & 2 \\
        0 & 0 & 0 & 0 & 0 & 1 & 0 & 1 & 0 \\
        0 & 0 & 0 & 0 & 0 & 0 & 1 & 0 & 0 \\
        0 & 0 & 0 & 0 & 0 & 0 & 0 & 1 & 0 \\
        0 & 0 & 0 & 0 & 0 & 0 & 0 & 0 & 1
  \end{matrix}\; \right) }
  \xrightarrow[\cx^2_{4,3}]{P_3:=P_3-P_4}
  \scalemath{\myscale}{\left(\; \begin{matrix}
        1 & 0 & 0 & 1 & 0 & 0 & 0 & 1 & 0 \\
        0 & 2 & 0 & 0 & 0 & 1 & 1 & 0 & 0 \\
        0 & 0 & 1 & 0 & 0 & 0 & 2 & 1 & 0 \\
        0 & 0 & 0 & 1 & 0 & 2 & 0 & 0 & 1 \\
        0 & 0 & 0 & 0 & 2 & 1 & 1 & 1 & 2 \\
        0 & 0 & 0 & 0 & 0 & 1 & 0 & 1 & 0 \\
        0 & 0 & 0 & 0 & 0 & 0 & 1 & 0 & 0 \\
        0 & 0 & 0 & 0 & 0 & 0 & 0 & 1 & 0 \\
        0 & 0 & 0 & 0 & 0 & 0 & 0 & 0 & 1
  \end{matrix}\; \right)}\notag\\
  \xrightarrow[\cx^2_{9,4}]{P_4:=P_4-P_9}
    &
    \scalemath{\myscale}{\left(\;\begin{matrix}
          1 & 0 & 0 & 1 & 0 & 0 & 0 & 1 & 0 \\
          0 & 2 & 0 & 0 & 0 & 1 & 1 & 0 & 0 \\
          0 & 0 & 1 & 0 & 0 & 0 & 2 & 1 & 0 \\
          0 & 0 & 0 & 1 & 0 & 2 & 0 & 0 & 0 \\
          0 & 0 & 0 & 0 & 2 & 1 & 1 & 1 & 2 \\
          0 & 0 & 0 & 0 & 0 & 1 & 0 & 1 & 0 \\
          0 & 0 & 0 & 0 & 0 & 0 & 1 & 0 & 0 \\
          0 & 0 & 0 & 0 & 0 & 0 & 0 & 1 & 0 \\
          0 & 0 & 0 & 0 & 0 & 0 & 0 & 0 & 1
    \end{matrix}\; \right)}
    \xrightarrow[\cx_{9,8}]{P_8:=P_8+P_9}
    \scalemath{\myscale}{
      \left(\;\begin{matrix}
          1 & 0 & 0 & 1 & 0 & 0 & 0 & 1 & 0 \\
          0 & 2 & 0 & 0 & 0 & 1 & 1 & 0 & 0 \\
          0 & 0 & 1 & 0 & 0 & 0 & 2 & 1 & 0 \\
          0 & 0 & 0 & 1 & 0 & 2 & 0 & 0 & 0 \\
          0 & 0 & 0 & 0 & 2 & 1 & 1 & 1 & 2 \\
          0 & 0 & 0 & 0 & 0 & 1 & 0 & 1 & 0 \\
          0 & 0 & 0 & 0 & 0 & 0 & 1 & 0 & 0 \\
          0 & 0 & 0 & 0 & 0 & 0 & 0 & 1 & 1 \\
          0 & 0 & 0 & 0 & 0 & 0 & 0 & 0 & 1
    \end{matrix}\; \right)}
    \xrightarrow[\cx_{8,5}]{P_5:=P_5+P_8}
    \scalemath{\myscale}{
      \left(\;\begin{matrix}
          1 & 0 & 0 & 1 & 0 & 0 & 0 & 1 & 0 \\
          0 & 2 & 0 & 0 & 0 & 1 & 1 & 0 & 0 \\
          0 & 0 & 1 & 0 & 0 & 0 & 2 & 1 & 0 \\
          0 & 0 & 0 & 1 & 0 & 2 & 0 & 0 & 0 \\
          0 & 0 & 0 & 0 & 2 & 1 & 1 & 2 & 0 \\
          0 & 0 & 0 & 0 & 0 & 1 & 0 & 1 & 0 \\
          0 & 0 & 0 & 0 & 0 & 0 & 1 & 0 & 0 \\
          0 & 0 & 0 & 0 & 0 & 0 & 0 & 1 & 1 \\
          0 & 0 & 0 & 0 & 0 & 0 & 0 & 0 & 1
    \end{matrix}\; \right)} \nonumber \\
    \xrightarrow[\cx^2_{9,8}]{P_8:=P_8-P_9} 
    &
    \scalemath{\myscale}{
      \left(\;\begin{matrix}
          1 & 0 & 0 & 1 & 0 & 0 & 0 & 1 & 0 \\
          0 & 2 & 0 & 0 & 0 & 1 & 1 & 0 & 0 \\
          0 & 0 & 1 & 0 & 0 & 0 & 2 & 1 & 0 \\
          0 & 0 & 0 & 1 & 0 & 2 & 0 & 0 & 0 \\
          0 & 0 & 0 & 0 & 2 & 1 & 1 & 2 & 0 \\
          0 & 0 & 0 & 0 & 0 & 1 & 0 & 1 & 0 \\
          0 & 0 & 0 & 0 & 0 & 0 & 1 & 0 & 0 \\
          0 & 0 & 0 & 0 & 0 & 0 & 0 & 1 & 0 \\
          0 & 0 & 0 & 0 & 0 & 0 & 0 & 0 & 1
    \end{matrix}\; \right)}
    \xrightarrow{\hspace{0.3cm}} \cdots \xrightarrow{\hspace{0.3cm}}
    \scalemath{\myscale}{
      \left(\;\begin{matrix}
          1 & 0 & 0 & 0 & 0 & 0 & 0 & 0 & 0 \\
          0 & 2 & 0 & 0 & 0 & 0 & 0 & 0 & 0 \\
          0 & 0 & 1 & 0 & 0 & 0 & 0 & 0 & 0 \\
          0 & 0 & 0 & 1 & 0 & 0 & 0 & 0 & 0 \\
          0 & 0 & 0 & 0 & 2 & 0 & 0 & 0 & 0 \\
          0 & 0 & 0 & 0 & 0 & 1 & 0 & 0 & 0 \\
          0 & 0 & 0 & 0 & 0 & 0 & 1 & 0 & 0 \\
          0 & 0 & 0 & 0 & 0 & 0 & 0 & 1 & 0 \\
          0 & 0 & 0 & 0 & 0 & 0 & 0 & 0 & 1
    \end{matrix}\; \right)}
\end{align*}
 \unskip
We proceed by computing the decreasing Steiner tree and performing the respective row operations for every column in decreasing order. Eventually, the ternary parity map becomes diagonal. 

As a final step, for each diagonal element with $P_{jj}=2$ we perform the row operation $P_{j}:=2P_{j}$. This operation corresponds to application of the $X^{(12)}$ gate on the $j$-th qutrit, as follows:
\begin{equation}
  P=
  \scalemath{\myscale}{
    \left(\;\begin{matrix}
        1 & 0 & 0 & 0 & 0 & 0 & 0 & 0 & 0 \\
        0 & 2 & 0 & 0 & 0 & 0 & 0 & 0 & 0 \\
        0 & 0 & 1 & 0 & 0 & 0 & 0 & 0 & 0 \\
        0 & 0 & 0 & 1 & 0 & 0 & 0 & 0 & 0 \\
        0 & 0 & 0 & 0 & 2 & 0 & 0 & 0 & 0 \\
        0 & 0 & 0 & 0 & 0 & 1 & 0 & 0 & 0 \\
        0 & 0 & 0 & 0 & 0 & 0 & 1 & 0 & 0 \\
        0 & 0 & 0 & 0 & 0 & 0 & 0 & 1 & 0 \\
        0 & 0 & 0 & 0 & 0 & 0 & 0 & 0 & 1
  \end{matrix}\; \right)}
  \xrightarrow[\scriptstyle X^{(12)}_2 X^{(12)}_5 ]{P_2:=2P_2,\; P_5:=2P_5}
  \scalemath{\myscale}{ \left(\;\begin{matrix}
        1 & 0 & 0 & 0 & 0 & 0 & 0 & 0 & 0 \\
        0 & 1 & 0 & 0 & 0 & 0 & 0 & 0 & 0 \\
        0 & 0 & 1 & 0 & 0 & 0 & 0 & 0 & 0 \\
        0 & 0 & 0 & 1 & 0 & 0 & 0 & 0 & 0 \\
        0 & 0 & 0 & 0 & 1 & 0 & 0 & 0 & 0 \\
        0 & 0 & 0 & 0 & 0 & 1 & 0 & 0 & 0 \\
        0 & 0 & 0 & 0 & 0 & 0 & 1 & 0 & 0 \\
        0 & 0 & 0 & 0 & 0 & 0 & 0 & 1 & 0 \\
        0 & 0 & 0 & 0 & 0 & 0 & 0 & 0 & 1
  \end{matrix}\; \right)} \notag
\end{equation}
Finally, the ternary parity map becomes an identity and the circuit $C^{-1}$ performs the same ternary string to ternary string mapping as $P$.

\section{Conclusion and outlook}

Recent progress in qutrit-based quantum computing makes it timely to develop circuit design tools, including gate compilation, for these systems. Motivated by this, we introduced a method to decompose any diagonal qutrit gate into \cx and single-qutrit gates, providing a direct analogue to the qubit case. To achieve this, we proposed a generalization of the Pauli gadget to qutrits, namely the Weyl gadget. This decomposition serves as a building block for qutrit-based quantum algorithms that require the implementation of such unitaries. As an additional result, we derived that implementing a generic diagonal qutrit unitary requires $O(3^{N})$ \cx gates. To illustrate the utility of our method, we applied the decomposition method to the quantum approximate optimization algorithm for the graph $k$-coloring problem for a number of colors $k\in\{3,9,27\}$, extending the results of \cite{bottrill2023}. The qutrit-based QAOA circuit exhibits advantages over its qubit-based counterpart, requiring fewer qudits and achieving lower circuit depth. This advantage becomes more pronounced as $k$ increases.

The proposed decomposition scheme has potential applications in other combinatorial optimization problems that could employ ternary encoding, such as the Traveling Salesman Problem (TSP) \cite{glos2020} and portfolio rebalancing \cite{huang_2024}. Looking for suitable ternary encoded Hamiltonians for these problems and comparing the resource cost, such as the number of qudits and circuit depth, against the qubit-counterpart, would be an interesting direction for future research.

Moreover, we extended the \steinergauss method to face connectivity limitations in qutrit systems by replacing the parity map with a ternary parity map on $\mathbb{Z}_3$ that accounts for the third computational state. This provides an intermediate representation for circuits consisting of CX and $X^{(12)}$ gates, that facilitates the mapping to topologies that do not allow direct communication among qutrits. The qutrit-based \steinergauss method improves the CX count compared to a trivial swap-based approach, yielding a reduction comparable to that achieved by its qubit counterpart. Further development would involve generalizing the approach to a qudit system. A natural starting point would be to replace the ternary parity map with a matrix over the field $\mathbb{Z}_d$ for prime $d$. Composite dimensions $d$ do not yield a field, and would require separate treatment.

By advancing qutrit-based gate decomposition methods and optimizing circuit layouts, this work contributes to the broader goal of improving the efficiency and scalability of qutrit quantum computing. Further exploration of qutrit-based algorithms and their hardware implementation may be crucial in determining the practical benefits of higher-dimensional quantum computation.

\section{Acknowledgments}

We would like to thank Arianne van de Griend and Samuel Whaite for useful discussions. We acknowledge funding from the Research Council of Finland through the Center of Excellence program Grant No. 336810 and through the Finnish Quantum Flagship projects 358878 (UH) and 358877 (Aalto). M.C. acknowledges funding from COQUSY Project No. PID2022-140506NB-C21 funded by Grant No. MCIN/AEI/10.13039/501100011033.

\bibliographystyle{quantum}
\bibliography{main}
\end{document}